\documentclass[twocolumn,showkeys,showpacs,preprintnumbers,prd,superscriptaddress,nofootinbib,aps,10pt]{revtex4-1}
\usepackage{graphicx,epsf,bm,amsmath,amsfonts,amssymb,epstopdf,natbib,verbatim,multirow,appendix,array,diagbox,xcolor,tikz,hyperref,booktabs,siunitx}
\usepackage[normalem]{ulem}

\hypersetup{colorlinks=true,urlcolor=blue,citecolor=blue,linkcolor=blue,menucolor=blue,anchorcolor=blue,filecolor=blue}

\newcolumntype{?}{!{\vrule width 3pt}}

\date{\today}

\makeatletter

\newcommand{\fmarki}{\ensuremath{\alpha}}
\newcommand{\fmarkii}{\ensuremath{\beta}}
\newcommand{\fmarkiii}{\ensuremath{\gamma}}
\newcommand{\fmarkiv}{\ensuremath{\delta}}
\newcommand{\fmarkv}{\ensuremath{\epsilon}}

\def\@fnsymbol#1{{\ifcase#1\or \fmarki\or \fmarkii\or \fmarkiii\or \fmarkiv\or \fmarkv\or \else\@ctrerr\fi}}
\makeatother
\definecolor{darkred}{rgb}{0.8, 0.0, 0.0}
\definecolor{darkgreen}{rgb}{0, 0.8, 0.0}
\definecolor{darkpink}{rgb}{0.8, 0.0, 0.4}

\definecolor{lime}{HTML}{A6CE39}
\DeclareRobustCommand{\orcidicon}{
\begin{tikzpicture}
\draw[lime, fill=lime] (0,0) 
circle [radius=0.16] 
node[white] {{\fontfamily{qag}\selectfont \tiny ID}};
\draw[white, fill=white] (-0.0625,0.095) 
circle [radius=0.007];
\end{tikzpicture}
\hspace{-3mm}
}

\foreach \x in {A, ..., Z}{\expandafter\xdef\csname orcid\x\endcsname{\noexpand\href{https://orcid.org/\csname orcidauthor\x\endcsname}{\noexpand\orcidicon}}}

\begin{document}

\title{Shape wall: Constraints on late-time solutions to the Hubble tension from the dimensionless expansion rate}

\author{\mbox{Miguel A. Sabogal\hspace{-1mm}\orcidC{}}}
\email{miguel.sabogalgarcia@unitn.it}
\affiliation{Department of Physics, University of Trento, Via Sommarive 14, 38123 Povo (TN), Italy}
\affiliation{Trento Institute for Fundamental Physics and Applications (TIFPA)-INFN, Via Sommarive 14, 38123 Povo (TN), Italy}

\author{\mbox{Luis A. Escamilla\hspace{-1mm}\orcidA{}}}
\email{torresl@itu.edu.tr}
\affiliation{Department of Physics, Istanbul Technical University, 34469 Maslak, Istanbul, Turkey \looseness=-1}

\author{\mbox{Davide Pedrotti\hspace{-1mm}\orcidB{}}}
\email{davide.pedrotti-1@unitn.it}
\affiliation{Department of Physics, University of Trento, Via Sommarive 14, 38123 Povo (TN), Italy}
\affiliation{Trento Institute for Fundamental Physics and Applications (TIFPA)-INFN, Via Sommarive 14, 38123 Povo (TN), Italy}

\author{\mbox{Edvig Selimaj\hspace{-1mm}\orcidD{}}}
\email{edvig.selimaj@studenti.unitn.it}
\affiliation{Department of Physics, University of Trento, Via Sommarive 14, 38123 Povo (TN), Italy}

\author{\mbox{Sunny Vagnozzi\hspace{-1mm}\orcidE{}}}
\email{sunny.vagnozzi@unitn.it}
\affiliation{Department of Physics, University of Trento, Via Sommarive 14, 38123 Povo (TN), Italy}
\affiliation{Trento Institute for Fundamental Physics and Applications (TIFPA)-INFN, Via Sommarive 14, 38123 Povo (TN), Italy}

\begin{abstract}
\noindent The standard ``no-go theorem'' against late-time solutions to the Hubble tension is essentially a normalization wall, since Baryon Acoustic Oscillation (BAO) measurements constrain the product $H_0r_d$, with $r_d$ the sound horizon at baryon drag. However, late-time solutions (which keep $r_d$ fixed) are tightly constrained not only by the BAO normalization $H_0r_d$, but also by the shape of the expansion history, i.e.\ the dimensionless expansion rate $E(z) \equiv H(z)/H_0$. We show that, if $r_d$ and the acoustic angular scale $\theta_s$ are fixed, an increase in $H_0$ needs to be matched by an equal fractional increase in the dimensionless distance integral $I \equiv \int dz/E(z)$: $\delta H_0/H_0 \simeq \delta I/I$. We use this to quantify the ``shape wall'' set by relative distance constraints on $E(z)$, which we reconstruct nonparametrically, using Gaussian Processes and the latest unanchored Type Ia Supernovae (SNeIa) and BAO data. In our most conservative analysis using \textit{PantheonPlus} SNeIa and \textit{DESI} DR2 BAO data, we find a maximum fractional increase in $H_0$ of $\lesssim 2\%$, falling well short of the $\gtrsim 8\%$ required to solve the tension. This shape wall holds even in the presence of early-time new physics, and limits the maximum increase in $H_0$ which can be contributed by late-time modifications to $E(z)$ at fixed $\theta_s$. Therefore, late-time modifications, whether invoked alone or alongside early-time new physics, face not only the well-known normalization wall, but also a stringent percent-level shape wall.
\end{abstract}

\maketitle

\section{Introduction}
\label{sec:introduction}

Among a number of cosmological tensions which have emerged over the past decade, the Hubble tension undoubtedly stands out as the most significant~\cite{Verde:2019ivm,DiValentino:2020zio,DiValentino:2021izs,Perivolaropoulos:2021jda,Shah:2021onj,Abdalla:2022yfr,DiValentino:2022fjm,Hu:2023jqc,Vagnozzi:2023nrq,Verde:2023lmm,CosmoVerseNetwork:2025alb,Cai:2026swf,Schoneberg:2026eys,Schoneberg:2026vaf,Wang:2026jxd}.~\footnote{The strength of the $S_8$ tension is milder, and depends significantly on the adopted dataset~\cite{DiValentino:2018gcu,DiValentino:2020vvd,Nunes:2021ipq,Wright:2025xka,Pantos:2026koc}.} For instance, the recent Local Distance Network (H0DN) estimate of $H_0=(73.50 \pm 0.81)\,{\text{km}}/{\text{s}}/{\text{Mpc}}$~\cite{H0DN:2025lyy} is in more than $7\sigma$ tension with the value inferred within $\Lambda$CDM from a combination of the latest \textit{Planck}, Atacama Cosmology Telescope (\textit{ACT}), and South Pole Telescope (\textit{SPT}) Cosmic Microwave Background (CMB) data, $H_0=(67.24 \pm 0.35)\,{\text{km}}/{\text{s}}/{\text{Mpc}}$~\cite{SPT-3G:2025bzu}. With explanations based on systematics facing increasing difficulties~\cite{Efstathiou:2020wxn,Mortsell:2021nzg,Mortsell:2021tcx,Freedman:2021ahq,Kenworthy:2022jdh,Camarena:2022iae,Wojtak:2022bct,Riess:2023bfx,Giani:2023aor,Wojtak:2024mgg,Freedman:2024eph,Perivolaropoulos:2024yxv,Riess:2025chq,Hogas:2026urs,Sorrenti:2026tye}, the persistence of the tension has made new physics beyond the $\Lambda$CDM model a very serious possibility (see e.g.\ Refs.~\cite{Karwal:2016vyq,Mortsell:2018mfj,Vagnozzi:2018jhn,Yang:2018euj,Poulin:2018dzj,RoyChoudhury:2018gay,Guo:2018ans,Poulin:2018cxd,Kreisch:2019yzn,Agrawal:2019lmo,Lin:2019qug,Yang:2019nhz,Vagnozzi:2019ezj,Visinelli:2019qqu,DiValentino:2019ffd,Smith:2019ihp,DiValentino:2019jae,Niedermann:2019olb,Berghaus:2019cls,Sakstein:2019fmf,Vagnozzi:2019kvw,Hart:2019dxi,Nojiri:2019fft,Ye:2020btb,Krishnan:2020obg,Zumalacarregui:2020cjh,Ballesteros:2020sik,Alestas:2020mvb,Jedamzik:2020krr,Braglia:2020iik,Ballardini:2020iws,DiValentino:2020evt,Gogoi:2020qif,Braglia:2020bym,Gonzalez:2020fdy,Sekiguchi:2020teg,Ye:2020oix,Niedermann:2020qbw,Murgia:2020ryi,Smith:2020rxx,CarrilloGonzalez:2020oac,Braglia:2020auw,Adi:2020qqf,Oikonomou:2020qah,Oikonomou:2020oex,RoyChoudhury:2020dmd,Brinckmann:2020bcn,Gao:2021xnk,Marra:2021fvf,SolaPeracaula:2021gxi,Dainotti:2021pqg,Teng:2021cvy,Krishnan:2021dyb,Vagnozzi:2021tjv,Vagnozzi:2021gjh,Karwal:2021vpk,Jiang:2021bab,Gomez-Valent:2021cbe,Hart:2021kad,Ye:2021iwa,Cyr-Racine:2021oal,Anchordoqui:2021gji,Akarsu:2021fol,Niedermann:2021ijp,Niedermann:2021vgd,Ferlito:2022mok,Wang:2022jpo,Nunes:2022bhn,Nojiri:2022ski,Oikonomou:2022yle,Schoneberg:2022grr,Reeves:2022aoi,RoyChoudhury:2022rva,Gomez-Valent:2022bku,Moshafi:2022mva,Rezazadeh:2022lsf,Escudero:2022rbq,Banerjee:2022ynv,deSa:2022hsh,Akarsu:2022typ,Lee:2022gzh,Khodadi:2023ezj,Bernui:2023byc,Ben-Dayan:2023rgt,Poulin:2023lkg,Cruz:2023lmn,Gomez-Valent:2023hov,Odintsov:2023cli,Ruchika:2023ugh,Adil:2023exv,Frion:2023xwq,Akarsu:2023mfb,Sharma:2023kzr,Gomez-Valent:2023uof,Efstathiou:2023fbn,Pedreira:2023qqt,Khalife:2023qbu,Akarsu:2024qiq,Erdem:2024vsr,Giare:2024ytc,Lynch:2024gmp,Giare:2024smz,Giare:2024akf,Lynch:2024hzh,Yadav:2024duq,Toda:2024ncp,Nozari:2024wir,Li:2024qso,Escamilla:2024xmz,Chatrchyan:2024xjj,RoyChoudhury:2024wri,Jiang:2024nha,Mirpoorian:2024fka,Gomez-Valent:2024ejh,Li:2025owk,Sabogal:2025mkp,Jiang:2025ylr,Stahl:2025czl,Akarsu:2025gwi,Jiang:2025hco,Yang:2025vnm,Lee:2025yah,Poulin:2025nfb,Yashiki:2025loj,Toda:2025kcq,Wang:2025dzn,Erdem:2025xtr,Zhang:2025dwu,Efstratiou:2025iqi,Chen:2025nfy,Kumar:2025obb,Hogas:2025mii,Pantos:2026cxv,Li:2026xaz,Bouhmadi-Lopez:2026dte,Dai:2026pvx,Wang:2026rkx,Gonzalez-Fuentes:2026rgu,Mukhopadhyay:2026fyk,Pantos:2026rpe,Bella:2026zuk,Carloni:2026yut,Pedrotti:2026dwj,Wang:2026kor,Jusufi:2026rfd,Dhyani:2026trw,Giare:2026tyk,Li:2026hwq,Akarsu:2026lva,Jia:2026vdt,Du:2026qtq,Li:2026asg,Hashim:2026yoy,Sabogal:2026qvy,Ajith:2026qqq,DOnofrio:2026dyq,Tsilioukas:2026gvy}). Any new physics solution must keep the acoustic angular scale $\theta_s$ fixed, as this quantity is inferred from CMB observations to extremely high precision and with very little model-dependence. The acoustic angular scale is given by $\theta_s=r_s^{\star}/D_A^{\star}$, with $r_s^{\star}$ being the comoving sound horizon at recombination, and $D_A^{\star}$ the comoving angular diameter distance to last-scattering. Broadly speaking, there are two qualitatively different routes to accommodate a larger inferred value of $H_0$ while keeping $\theta_s$ fixed. Early-time attempts modify the pre-recombination expansion history to reduce $r_s^{\star}$, with $D_A^{\star}$ then decreasing by the same amount. Late-time attempts instead leave $r_s^{\star}$ unchanged, but modify the ``shape'' of the post-recombination expansion rate, i.e.\ the dimensionless expansion rate $E(z) \equiv H(z)/H_0$, in order to increase $H_0$ while keeping $D_A^{\star}$ unchanged. At the geometrical level, CMB measurements do not on their own distinguish between these two classes of solutions.

A key role in breaking this geometrical degeneracy is played by Baryon Acoustic Oscillation (BAO) data, which provide measurements of low-redshift distances relative to the comoving sound horizon at baryon drag, $r_d$, a quantity closely related to $r_s^{\star}$ by $r_d \simeq 1.018 r_s^{\star}$. This makes BAO measurements sensitive to the combination $r_dH_0$, and therefore, once $r_d$ is appropriately calibrated, to the absolute normalization of the late-time expansion history. Numerically, current BAO data are broadly consistent with $r_dH_0\simeq 10^4\,{\text{km}}/{\text{s}}$, which within $\Lambda$CDM corresponds to a high sound horizon, $r_d \simeq 147\,{\text{Mpc}}$, and a low Hubble constant, $H_0 \simeq 67\,{\text{km}}/{\text{s}}/{\text{Mpc}}$. Raising $H_0$ while preserving this normalization requires a proportional relative decrease in $r_d$, i.e.\ $\delta r_d/r_d \simeq -\delta H_0/H_0$. Because post-recombination modifications alone are unable to change $r_d$, they cannot raise $H_0$ while maintaining agreement with BAO data. This is the key argument of the well-known ``no-go theorem'' precluding a purely late-time solution to the Hubble tension~\cite{Bernal:2016gxb,Addison:2017fdm,Lemos:2018smw,Aylor:2018drw,Schoneberg:2019wmt,Knox:2019rjx,Arendse:2019hev,Efstathiou:2021ocp,Cai:2021weh,Keeley:2022ojz}. At its core, this is therefore essentially a \textit{normalization} argument, making the main obstruction to late-time solutions a ``normalization wall''.

However, the normalization wall is not necessarily the end of the story, as it is known that the shape of the expansion history is also tightly constrained, for instance by unanchored Type Ia Supernovae (SNeIa), see e.g.\ Refs.~\cite{Bernal:2016gxb,Jiang:2024xnu,Zhou:2025kws,Pedrotti:2025ccw,Zhou:2026iar}. Recent work has in fact argued that considerations on the dimensionless expansion rate $E(z)$ can play a particularly important role in the Hubble tension. For instance, even before taking BAO data into account, relative distance measurements from unanchored SNeIa alone have recently been shown to place very tight constraints on late-time attempts to solve the Hubble tension~\cite{Camarena:2021jlr,Zhou:2025kws,Pedrotti:2025ccw,Bansal:2026axl,Zhou:2026iar}. These considerations motivate the central question of our work: regardless of the absolute normalization (or after marginalizing over it), how much freedom is actually left in the shape of the late-time expansion history $E(z)$? Stated differently, is there a ``shape wall'' independent of the usual $r_dH_0$ normalization wall, and if so how strong is it?

It is our goal in this work to quantify this shape wall. Our key observation is that, once $\theta_s$ is fixed, the allowed relative change in $H_0$ can be related to the relative change in the integral of $1/E(z)$. In order to be as model-agnostic as possible, we reconstruct $E(z)$ nonparametrically from late-time observations in a way that is insensitive to the absolute normalization, thereby isolating shape information. We then propagate the reconstructed dimensionless expansion histories $E(z)$ to determine the relative shift in $H_0$, $\delta H_0/H_0$, allowed by late-time shape information alone. We find that, once unanchored SNeIa and BAO data are combined, the allowed shifts are at the percent level at best, far below the $\simeq 8$-$9\%$ change required to fully solve the Hubble tension without modifying the sound horizon. This confirms that a percent-level shape wall exists, and that the obstruction to late-time solutions is not only one of normalization, or calibration, but also one of shape. This distinction is important because, unlike a normalization wall, evading a shape wall requires more than a shift in the absolute calibration: it calls for redshift-dependent changes in the observed distance-redshift relation (see e.g.\ Refs.~\cite{Renzi:2021xii,Teixeira:2025czm,Li:2025htp,Zhou:2025dxo,Tiwari:2026pzk}), whether due to new physics or systematics.

The rest of this work is then organized as follows. In Sec.~\ref{sec:theoretical} we discuss our theoretical framework for connecting the reconstructed dimensionless expansion history $E(z)$ to the allowed relative shift in $H_0$. In Sec.~\ref{sec:datasets} we discuss the datasets and methodology adopted to obtain our results. These are presented in Sec.~\ref{sec:results}, before being critically discussed in Sec.~\ref{sec:discussion}. Finally, in Sec.~\ref{sec:conclusions} we draw concluding remarks. A detailed discussion of more technical aspects of our nonparametric reconstruction methodology is presented in Appendix~\ref{app:gpmethodology}, whereas the results obtained from an alternative node-based reconstruction methodology are shown in Appendix~\ref{app:alternativegpmethodology}.

\section{Theoretical framework}
\label{sec:theoretical}

We begin by briefly recalling the normalization wall ``no-go theorem'' against purely late-time solutions set by BAO measurements. Throughout this work, we assume a homogeneous and isotropic cosmology described by the spatially flat Friedmann-Lema\^{i}tre-Robertson-Walker (FLRW) metric. The transverse comoving distance is then determined as follows:
\begin{equation}
D_M(z)=\frac{1}{H_0}\int_0^z\,\frac{dz'}{E(z')}\,,
\label{eq:dm}
\end{equation}
where $H_0$ is the Hubble constant, $E(z)$ is the dimensionless expansion rate, and we are implicitly setting $c=1$. Transverse, line-of-sight, and volume-averaged BAO measurements at an effective redshift $z_{\text{eff}}$ are then sensitive to the transverse angular scale $\theta_d$, redshift span $\delta z_d$, and volume-averaged angular scale $\theta_v$, which are given by the following:
\begin{align}
\label{eq:thetad}
\theta_d(z_{\text{eff}}) &= \frac{r_d}{D_M(z_{\text{eff}})}=\frac{r_dH_0}{\displaystyle \int_0^{z_{\text{eff}}}\frac{dz'}{E(z')}}\,, \\
\label{eq:zd}
\delta z_d(z_{\text{eff}}) &= \frac{r_d}{D_H(z_{\text{eff}})} \equiv r_d H(z_{\text{eff}})= r_dH_0 E(z_{\text{eff}})\,,\\
\label{eq:thetav}
\theta_{V}(z_{\text{eff}}) &= \frac{r_d}{D_V(z_{\text{eff}})} \equiv \frac{r_d}{ \left [ z_{\text{eff}} D_M^2(z_{\text{eff}}) D_H(z_{\text{eff}}) \right ] ^{1/3}}\,,
\end{align}
where the comoving size of the sound horizon at baryon drag $r_d$ plays the role of standard ruler. From Eqs.~(\ref{eq:thetad}--\ref{eq:thetav}) it is clear that BAO measurements are sensitive to the combination $r_dH_0$, which controls the angular size of the sound horizon. Any new physics model which raises $H_0$ must therefore do so while decreasing $r_d$, or risk running into disagreement with BAO data. As stressed earlier, this essentially sets a normalization wall for the viability of late-time approaches to the Hubble tension. By construction, these models are unable to change $r_d$, as this quantity is determined by an integral running over redshifts $z>z_{\text{drag}}$, with $z_{\text{drag}} \simeq 1060$ being the redshift of baryon drag. This is the reason behind the common expectation that resolving the Hubble tension requires new pre-recombination physics~\cite{Bernal:2016gxb,Addison:2017fdm,Lemos:2018smw,Aylor:2018drw,Schoneberg:2019wmt,Knox:2019rjx,Arendse:2019hev,Efstathiou:2021ocp,Cai:2021weh,Keeley:2022ojz}.

Late-time distance measurements, however, contain more information than their absolute normalization alone. In fact, recent work has argued that the shape of the expansion history, i.e.\ the dimensionless expansion rate $E(z)$, plays a particularly important role in constraining late-time attempts to address the Hubble tension~\cite{Camarena:2021jlr,Zhou:2025kws,Pedrotti:2025ccw,Bansal:2026axl,Zhou:2026iar}. The reason is that these models aim to increase $H_0$ while keeping $r_s^{\star}$ fixed. In order to keep $D_A^{\star}$ fixed, the larger value of $H_0$ needs to be compensated by a change in $E(z)$ or, more precisely, in the integral of $1/E(z)$. Any constraint on the shape of the late-time expansion history $E(z)$, typically arising from relative cosmological distances, will therefore set limits on the viability of late-time attempts to solve the Hubble tension. This is the motivation behind the main question our work seeks to answer: once information about the absolute normalization is removed, or explicitly marginalized over, how much freedom remains for these models in the shape of $E(z)$?

\begin{figure*}[!htpb]
\centering
\includegraphics[width=0.75\linewidth]{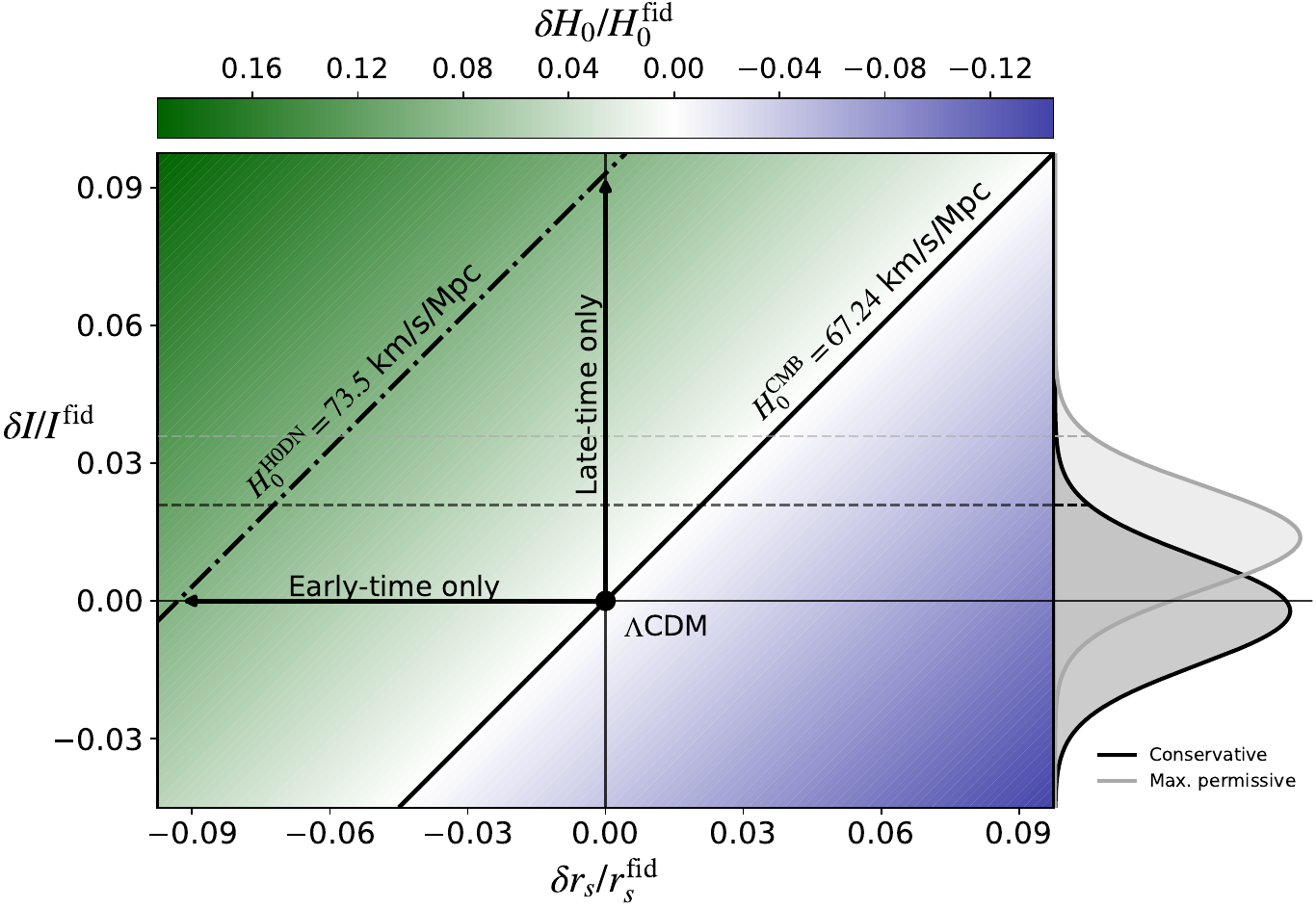}
\caption{Visual representation of the shape wall. We show the geometrical conditions required to shift the Hubble constant $H_0$, when fixing the acoustic angular scale $\theta_s$ [see Eq.~(\ref{eq:deltathetas}), with $\delta\theta_s/\theta_s \approx 0$], with green (purple) indicating larger (smaller) values of $H_0$, and the black dot indicating the fiducial $\Lambda$CDM model. Purely late-time and purely early-time modifications corresponding to moving along the vertical and horizontal directions respectively. The solid (dash-dotted) diagonal corresponds to the value of $H_0$ inferred from \textit{Planck}+\textit{ACT}+\textit{SPT} data within $\Lambda$CDM (the Local Distance Network), with the region in between the two being the one where the Hubble tension is alleviated. On the right we show the posterior distributions for $\delta I/I^{\text{fid}}$ within two benchmark scenarios to be discussed in Sec.~\ref{sec:discussion} (see Tab.~\ref{tab:deltaipercentage}): a more conservative scenario (black) and a maximally permissive scenario (gray), with dashed lines indicating the respective 95\% upper limits reported in Tab.~\ref{tab:deltaipercentage}. This figure is based on a related figure originally presented by William Giar\`{e}, whom we thank for sharing his plotting code, publicly available at \href{https://github.com/williamgiare/wgcosmo/blob/main/plots/Hubble\_Tension/Mixed\_Solutions.ipynb}{https://github.com/williamgiare/wgcosmo/blob/main/plots/Hubble\_Tension/Mixed\_Solutions.ipynb}, which we have adapted.}
\label{fig:summary}
\end{figure*}

To make this statement quantitative, let us consider the acoustic angular scale, which we rewrite as follows:
\begin{equation}
\theta_s=\frac{r_s^{\star}}{D_A^{\star}}=\frac{r_s^{\star}H_0}{I}\,,
\label{eq:thetas}
\end{equation}
where $I$ (which stands for ``integral'') is defined as follows:
\begin{equation}
I \equiv \int_{0}^{z_{\star}}\,\frac{dz}{E(z)}
\label{eq:integral}
\end{equation}
with $z_{\star} \simeq 1090$ being the redshift of the last-scattering surface. Considering small relative variations, we can perturb Eq.~(\ref{eq:thetas}) as follows:
\begin{equation}
\frac{\delta\theta_s}{\theta_s} \simeq \frac{\delta r_s^{\star}}{r_s^{\star}}+\frac{\delta H_0}{H_0}-\frac{\delta I}{I}\,,
\label{eq:deltathetas}
\end{equation}
where the variations $\delta\theta_s$, $\delta r_s^{\star}$, $\delta H_0$, and $\delta I$ are to be understood with respect to a reference ``fiducial'' model. Since $\theta_s$ is fixed to extremely high precision and with very little model-dependence from CMB observations, we require $\delta \theta_s/\theta_s \simeq 0$ in Eq.~(\ref{eq:deltathetas}). Moreover, a purely late-time solution to the Hubble tension cannot change the sound horizon, so $\delta r_s^{\star}/r_s^{\star}=0$ as well (in passing, we note that this implies $\delta r_d/r_d=0$). Therefore, for this class of models, Eq.~(\ref{eq:deltathetas}) reduces to:
\begin{equation}
\frac{\delta H_0}{H_0} \simeq \frac{\delta I}{I}\,.
\label{eq:deltah0}
\end{equation}
The above equation sets in mathematical terms a statement we made in the Introduction: the allowed relative change in $H_0$ is related to (and, more precisely, equal to) the relative change in the integral of $1/E(z)$. Thus, if the sound horizon is fixed, raising $H_0$ requires increasing $I$ by the same relative amount (see also Ref.~\cite{Giare:2026tyk}).

The above equations are visually represented in Fig.~\ref{fig:summary}, which shows the geometrical conditions required to shift $H_0$ at fixed $\theta_s$, i.e.\ Eq.~(\ref{eq:deltathetas}) with $\delta\theta_s/\theta_s \approx 0$. Starting from the black dot corresponding to the fiducial $\Lambda$CDM model, purely late-time and purely early-time modifications correspond to moving along the vertical and horizontal directions respectively. Mixed (early-plus-late) modifications instead fall in between: the more the late-time part contributes, the more the corresponding line rotates clockwise. The region where the Hubble tension is alleviated corresponds to that between the two diagonal (solid and dash-dotted) lines. On the right we show the posterior distributions for $\delta I/I^{\text{fid}}$ within two benchmark scenarios to be discussed in Sec.~\ref{sec:discussion} (see Tab.~\ref{tab:deltaipercentage}): a more conservative scenario (black) and a maximally permissive scenario (gray), with dashed lines indicating the respective 95\% upper limits reported in Tab.~\ref{tab:deltaipercentage}.

The size of the required effect for a complete resolution of the Hubble tension is easy to estimate. For concreteness, we compare the recent H0DN estimate of $H_0=(73.50 \pm 0.81)\,{\text{km}}/{\text{s}}/{\text{Mpc}}$~\cite{H0DN:2025lyy} against the value inferred within the $\Lambda$CDM model from a combination of the latest \textit{Planck}, \textit{ACT}, and \textit{SPT} CMB data, $H_0=(67.24 \pm 0.35)\,{\text{km}}/{\text{s}}/{\text{Mpc}}$~\cite{SPT-3G:2025bzu}. The two measurements are in $\gtrsim 7\sigma$ disagreement, and reconciling them requires a relative shift of $\delta H_0/H_0 \simeq 0.093$, i.e.\ nearly a $10\%$ increase in $H_0$. As is clear from Eqs.~(\ref{eq:deltathetas},\ref{eq:deltah0}), at fixed acoustic angular scale $\theta_s$ and sound horizon, a purely late-time model which fully solves the Hubble tension must generate an ${\cal O}(10\%)$ increase in the dimensionless distance integral $I$. Smaller values of $\delta I/I$, while still reducing the tension, would lead to a smaller increase in $H_0$. One may of course choose a more modest target and accept a residual tension: for example, the often-quoted $7\%$ increase in $H_0$, or equivalently $7\%$ decrease in $r_d$ for early-time solutions, would leave a residual Gaussian tension of $\simeq 1.8\sigma$, far from unacceptable. Our goal in this work is therefore to estimate the highest possible value of $\delta I/I$ allowed by current observations. This can then be directly propagated to $H_0$ using Eq.~(\ref{eq:deltah0}).

We see from Eq.~(\ref{eq:integral}) that the integral defining $I$ runs up to $z_{\star}$. However, existing late-time observations sensitive to $E(z)$ only cover a finite redshift range, indicatively $z<z_{\max} \approx 2.3$, with the precise value of $z_{\max}$ depending on the SNeIa and BAO samples utilized. For this reason, it is useful to split the integral into a part which can be directly probed by late-time data, and a tail at higher redshift which cannot be directly constrained. These contributions, which we refer to as $I_<$ and $I_>$ respectively, are defined as follows:
\begin{equation}
I_< \equiv \int_{0}^{z_{\max}}\,\frac{dz}{E(z)}\,, \qquad I_> \equiv \int_{z_{\max}}^{z_{\star}}\,\frac{dz}{E(z)}\,,
\label{eq:ilig}
\end{equation}
and obviously $I=I_<+I_>$. From late-time observations, we can then derive $\delta I_</I_<^{\text{fid}}$, where we introduce the superscript ``fid'' to stress that all variations are computed with respect to a given specified fiducial model. As we shall argue later, although the absence of data prevents a direct nonparametric estimate of $\delta I_>/I_>^{\text{fid}}$, it is actually possible to provide a reasonable model-guided estimate of this quantity. It is important to note that the relative shift in $I$, i.e.\ $\delta I/I^{\text{fid}}$, is \textit{not} simply the sum of $\delta I_</I_<^{\text{fid}}$ and $\delta I_>/I_>^{\text{fid}}$. To see why, note that we can express the relative variation of the total integral as follows:
\begin{equation}
\frac{\delta I}{I^{\text{fid}}}=\frac{\delta I_<+\delta I_>}{I^{\text{fid}}}=f_<\frac{\delta I_<}{I_<^{\text{fid}}}+f_>\frac{\delta I_>}{I_>^{\text{fid}}}\,,
\label{eq:deltaildeltaig}
\end{equation}
where the weights $f_<$ and $f_>$ are defined as follows:
\begin{equation}
f_< \equiv \frac{I_<^{\text{fid}}}{I^{\text{fid}}}\,, \qquad f_> \equiv \frac{I_>^{\text{fid}}}{I^{\text{fid}}}\,.
\label{eq:weights}
\end{equation}
Therefore, $\delta I/I^{\text{fid}}$ is actually given by a \textit{weighted} sum of $\delta I_</I_<^{\text{fid}}$ and $\delta I_>/I_>^{\text{fid}}$. For reference, within a fiducial $\Lambda$CDM model with $\Omega_m=0.31$, i.e.\ the best-fit \textit{Planck} $\Lambda$CDM cosmology which we will adopt, and $z_{\max}=2.3$ (approximately the maximum redshift probed by the \textit{DESI} DR2 BAO sample~\cite{DESI:2025zgx}) we find $f_< \simeq 0.41$ and $f_> \simeq 0.59$. In other words, roughly $41\%$ of the $I$ integral is contributed by redshifts $z \lesssim z_{\max}$, whereas the $z \gtrsim z_{\max}$ contribution is about $59\%$.

Having discussed our theoretical framework, the strategy of our work is then conceptually simple. Using Gaussian Processes (GPs)~\cite{Seikel:2012uu}, we nonparametrically reconstruct the dimensionless expansion history $E(z)$ for $z<z_{\max}$, using unanchored SNeIa data, alone and/or in combination with BAO measurements, in a way which is insensitive to the absolute calibration of the latter, and therefore isolates purely shape information. We then propagate these reconstructions to the full posterior of $\delta I_</I_<^{\text{fid}}$. We note that, at leading order around the fiducial model, this quantity can be approximated as follows:
\begin{equation}
\frac{\delta I_<}{I_<^{\text{fid}}} \simeq -\frac{1}{I_<^{\text{fid}}}\int_0^{z_{\max}}\,\frac{dz}{E_{\text{fid}}(z)}\frac{\delta E(z)}{E_{\text{fid}}(z)}\,.
\label{eq:deltaillinear}
\end{equation}
We stress that we do not rely on the above approximate expression, but evaluate $I_<$ directly for each reconstructed $E(z)$ to construct the full posterior distribution for $\delta I_</I_<^{\text{fid}}$. This is the main result of our analysis, i.e.\ a shape-only, nonparametric constraint on the part of the dimensionless distance integral $I$ directly probed by late-time observations, which can be mapped onto the maximum allowed shift in $H_0$ through Eqs.~(\ref{eq:deltah0},\ref{eq:deltaildeltaig}). As alluded to earlier, we can actually provide a reasonable model-guided estimate of $\delta I_>/I_>^{\text{fid}}$ despite the absence of data, of course not at the same level of robustness of our estimate of $\delta I_</I_<^{\text{fid}}$. Our strategy for estimating $\delta I_>/I_>^{\text{fid}}$, and therefore the total $\delta I/I^{\text{fid}}$ through Eq.~(\ref{eq:deltaildeltaig}), will be discussed in Sec.~\ref{sec:discussion}. We stress that, even if early-time new physics modifies $r_s^{\star}$, our results will still constrain the late-time contribution $(\delta H_0/H_0)_{\text{late}}=\delta I/I$, with $H_0$ receiving an additional, independent increase $(\delta H_0/H_0)_{\text{early}}=-\delta r_s^\star/r_s^\star$ from early-time new physics, as in Eq.~(\ref{eq:deltathetas}). In this case, the shape wall we will quantify will generally determine the maximum relative increase in $H_0$ which can be contributed by late-time modifications to $E(z)$ at fixed $\theta_s$, regardless of whether new physics is present at early times as well.

\section{Datasets and methodology}
\label{sec:datasets}

We now briefly discuss the datasets and methodology adopted in deriving our results. We remind the reader that a more detailed, technical discussion of our GP reconstruction procedure is presented in Appendix~\ref{app:gpmethodology}.

\subsection{Datasets}
\label{subsec:datasets}

To nonparametrically reconstruct $E(z)$ and determine our posteriors for $\delta I_</I_<^{\text{fid}}$ (and ultimately the allowed values of $\delta H_0/H_0$), we make use of the following late-time cosmological observations:
\begin{itemize}
\item SNeIa distance moduli measurements from the \textit{Pantheon+} sample, consisting of 1701 light curves for 1550 unique spectroscopically classified SNeIa~\cite{Brout:2022vxf}. We only use points in the redshift range $0.01<z<2.26$, therefore excluding the low-redshift SH0ES calibrator sample and ensuring that our SNeIa are used exclusively as relative (as opposed to absolute) distance indicators, which are only sensitive to $E(z)$. We refer to these measurements as \textit{PantheonPlus}.
\item SNeIa distance moduli measurements from the Dark Energy Survey Year 5 sample, recalibrated through the \textit{Dovekie} cross-calibration program~\cite{DES:2025sig}. This consists of 1623 photometrically classified SNeIa in the redshift range $0.1<z<1.13$, alongside a set of 197 SNeIa in the redshift range $0.025<z<0.1$ drawn from historical sources. We refer to these measurements as \textit{DESY5Dvk}.
\item SNeIa distance moduli measurements from the Union3 sample, consisting of 2087 spectroscopically classified SNeIa, binned in 22 bins in the redshift range $0.05<z<2.26$~\cite{Rubin:2023jdq}. We refer to these measurements as \textit{Union3}.
\item BAO measurements from the Dark Energy Spectroscopic Instrument (\textit{DESI}) DR2 release. The full dataset includes measurements of $D_V/r_{\text{drag}}$, $D_M/r_{\text{drag}}$, and $D_H/r_{\text{drag}}$ [see Eqs.~(\ref{eq:thetad}--\ref{eq:thetav}) for the definitions of these quantities] across 7 redshift bins in the range $0.295<z_{\text{eff}}<2.330$~\cite{DESI:2025zgx}. However, in our GP reconstruction we only adopt measurements of $D_M/r_{\text{drag}}$ and $D_H/r_{\text{drag}}$ across 6 redshift bins in the range $0.510<z_{\text{eff}}<2.330$ (since our choice leads us to exclude the BGS sample at $z_{\text{eff}}=0.295$), taking into account the covariance between the measurements.~\footnote{We do not include volume-averaged BAO measurements in our GP reconstruction, as these combine transverse and line-of-sight information into a nonlinear distance which does not directly match to the reconstruction we perform, and would require a modified likelihood treatment, as opposed to the direct reconstruction of $D_H$ and $D_M$ we perform. We refer the reader to Appendix~\ref{app:gpmethodology} for more details.} We refer to the combinations of these measurements as \textit{DESI}.
\end{itemize}
In our analysis we make use of six different combinations of the above datasets, three of which consider unanchored SNeIa alone, while the remaining three combine these with BAO measurements. Specifically, the six dataset combinations we consider are the following:
\begin{itemize}
\item \textit{PantheonPlus}
\item \textit{DESY5Dvk}
\item \textit{Union3}
\item \textit{PantheonPlus}+\textit{DESI}
\item \textit{DESY5Dvk}+\textit{DESI}
\item \textit{Union3}+\textit{DESI}
\end{itemize}
We stress that the value of $z_{\max}$ relevant for splitting the $I_<$ contribution from the $I_>$ one differs for each dataset combination considered. For the dataset combinations involving \textit{DESI}, we set $z_{\max}=2.33$, corresponding to the effective redshift of the highest redshift BAO measurements from the Lyman-$\alpha$ forest sample (Ly$\alpha$). For the \textit{PantheonPlus}, \textit{DESY5Dvk}, and \textit{Union3} datasets, we instead set $z_{\max}=2.26$, $1.13$, and $2.26$ respectively.\\

\subsection{Dimensionless distances}
\label{subsec:dimensionless}

Since we are interested in isolating shape information, it is useful to think in terms of dimensionless distances, i.e.\ distances multiplied by $H_0$.~\footnote{Recall that we are working in units where $c=1$, so combinations of the type $H_0D$, where $D$ is a distance, are dimensionless. In SI units, the relevant dimensionless combination is $H_0D/c$.} We define the dimensionless transverse comoving distance $\widetilde D_M(z)$ as follows:
\begin{equation}
\widetilde D_M(z) \equiv H_0D_M(z)=\int_0^z \frac{dz'}{E(z')}\,.
\label{eq:dmtilde}
\end{equation}
Taking the derivative of $\widetilde D_M(z)$ directly gives us the inverse of the dimensionless expansion rate, i.e.\ the dimensionless Hubble distance $\widetilde D_H(z)$:
\begin{equation}
\widetilde D_M'(z) \equiv \frac{d\widetilde D_M}{dz}=\frac{1}{E(z)} \equiv \widetilde D_H(z)\,.
\label{eq:dhtilde}
\end{equation}
Working with these dimensionless quantities makes it possible to isolate shape information. Our unanchored SNeIa measurements constrain the relative distance-redshift relation, allowing us to ultimately reconstruct $\widetilde D_M$. Our treatment of BAO measurements is discussed in more detail in Appendix~\ref{app:gpmethodology}. In a nutshell, we effectively turn BAO measurements of $D_M/r_{\rm drag}$ and $D_H/r_{\rm drag}$ into constraints on $\widetilde D_M$ and $\widetilde D_H$ by marginalizing over a common amplitude proportional to $H_0r_{\text{drag}}$. This makes our results sensitive not to the overall BAO normalization, but only to the redshift dependence of BAO distances, and the consistency between $\widetilde D_M$ and its derivative $\widetilde D_M'(z)$. In other words, we use BAO as a relative (as opposed to absolute) distance indicator. See Ref.~\cite{Zhang:2025bmk} for a recent work adopting a similar methodology.

From our GP reconstruction, we obtain an ensemble of correlated, reconstructed realizations of $\widetilde D_M'(z)$, from which we construct the corresponding dimensionless expansion histories through $E(z)=1/\widetilde D_M'(z)$. For each reconstructed $E(z)$, we then compute $I_<$ by integrating $1/E(z)$ up to the value of $z_{\max}$ appropriate for the relevant dataset combination. We then compare this to the fiducial value $I_<^{\text{fid}}$, obtained assuming a fiducial $\Lambda$CDM cosmology with $\Omega_m^{\text{fid}}=0.31$, as informed by \textit{Planck} CMB observations:
\begin{equation}
I_<^{\text{fid}}=\int_0^{z_{\max}}\frac{dz}{E_{\text{fid}}(z)}\,,
\label{eq:ilfid}
\end{equation}
where the fiducial dimensionless expansion history $E(z)^{\text{fid}}$ is given by the following expression:
\begin{equation}
E(z)^{\text{fid}} \simeq \sqrt{\Omega_m^{\text{fid}}(1+z)^3+(1-\Omega_m^{\text{fid}})}\,,
\label{eq:ezfid}
\end{equation}
with the $\simeq$ sign reflecting the fact that we have neglected the radiation component. For each realization of $E(z)$ within the GP-reconstructed ensemble, we finally derive the relative variation of $I_<^{\text{fid}}$ as follows:
\begin{equation}
\frac{\delta I_<}{I_<^{\text{fid}}} \equiv \frac{I_<-I_<^{\text{fid}}}{I_<^{\text{fid}}}\,.
\label{eq:deltail}
\end{equation}
Therefore, the posterior distribution for $\delta I_</I_<^{\text{fid}}$ is directly obtained from our ensemble of $E(z)$ reconstructions [as opposed to using a linearized approximation such as that of Eq.~(\ref{eq:deltaillinear})]. The approach we adopt allows us to consistently propagate uncertainties in the GP reconstruction (including those on the GP hyperparameters), and effectively amounts to treating $\delta I_</I_<^{\text{fid}}$ as a derived parameter. Within the assumed fiducial cosmology, we find $I^{\text{fid}} \approx 3.13$. The values of $I_<^{\text{fid}}$, $I_>^{\text{fid}}$, and thus the values of the weights $f_<$ and $f_>$ appearing in Eq.~(\ref{eq:weights}), depend on the specific value of $z_{\max}$, and therefore on the dataset combination adopted. For $z_{\max}=1.13$, $2.26$, and $2.33$, we find $I_<^{\text{fid}}=0.83$, $1.27$, and $1.30$ respectively, and $I_>^{\text{fid}}=2.30$, $1.86$, and $1.83$ respectively. The corresponding weights are therefore $f_<=0.27$, $0.41$, and $0.42$ respectively, and $f_>=0.73$, $0.59$, and $0.58$ respectively.

\subsubsection{Gaussian Process reconstruction}
\label{subsubsec:gpreconstruction}

We now briefly summarize our GP reconstruction methodology, which we use to obtain the ensemble of dimensionless expansion histories discussed above. Full technical details are provided in Appendix~\ref{app:gpmethodology}, which we encourage the interested reader to consult. We recall that GPs are a generalization of the concept of Gaussian distributions to the (infinite-dimensional) function space, and have found widespread use in cosmology in recent years (see e.g.\ Refs.~\cite{Holsclaw:2010sk,Shafieloo:2012ht,Zhang:2018gjb,Gerardi:2019obr,LHuillier:2019imn,Cai:2019bdh,Liao:2019qoc,Aljaf:2020eqh,Benisty:2020kdt,Bengaly:2020vly,Briffa:2020qli,Keeley:2020aym,Renzi:2020bvl,Bonilla:2020wbn,vonMarttens:2020apn,Bonilla:2021dql,Bora:2021bui,Bora:2021cjl,Mukherjee:2021kcu,Sun:2021pbu,Escamilla-Rivera:2021rbe,Bernardo:2021qhu,Bernardo:2021mfs,Bengaly:2021wgc,Jesus:2021bxq,Ruiz-Zapatero:2022zpx,Avila:2022xad,Benisty:2022psx,Mukherjee:2022ujw,Ren:2022aeo,Hwang:2022hla,Mukherjee:2023lqr,Liu:2023ifz,Liu:2023ulr,Dinda:2023xqx,Yang:2024tkw,Zhang:2024ndc,Mukherjee:2024ryz,Sabogal:2024qxs,Ghosh:2024kyd,Velazquez:2024aya,Mukherjee:2024pcg,Yang:2025kgc,vonMarttens:2025dvv,Ruchika:2025sbb,Abedin:2025dis,Li:2025ula,Cheng:2025cmb,Liu:2025evk,Avila:2025sjz,Rana:2025psr,Ruchika:2025mkx,Akarsu:2026anp}).

We begin by converting our SNeIa measurements into dimensionless comoving distances $\widetilde D_M(z)$, on a common reference scale (see also Ref.~\cite{Zhang:2025bmk}). In practice, for the \textit{PantheonPlus} sample we make use of the distance moduli $\mu_i=m_{\text{obs},i}-M^{\text{fid}}$, where $m_{\text{obs}}$ is the usual standardized apparent $B$-band peak magnitude, and $M^{\text{fid}}=-19.25$ is the fiducial absolute magnitude, associated to a reference fiducial value of $H_0^{\text{fid}}\simeq73\,{\text{km}}/{\text{s}}/{\text{Mpc}}$. We stress that we are \textit{not} adopting the SH0ES calibrator likelihood or imposing a prior on $M$ or $H_0$: rather, we are fixing $M^{\text{fid}}$ (or equivalently $H_0^{\text{fid}}$) in order to convert the \textit{PantheonPlus} apparent magnitudes into dimensionless distances using the same reference normalization as the \textit{Union3} and \textit{DESY5Dvk} SNeIa samples, as well as the \textit{DESI} BAO measurements. The actual value of $H_0^{\text{fid}}$ is arbitrary and will cancel in the later analysis. As for the \textit{Union3} and \textit{DESY5Dvk} samples, the publicly released distance moduli are conventionally ``normalized'' to $H_0^{\text{ref}}=70\,{\text{km}}/{\text{s}}/{\text{Mpc}}$~\cite{Rubin:2023jdq,DES:2025sig}. In order to adopt the same common reference scale as \textit{PantheonPlus}, we therefore shift these measurements as $\mu_i=\mu_{\text{obs},i}-5\log_{10}(H_0^{\text{fid}}/H_0^{\text{ref}})$, where $\mu_{\text{obs},i}$ are the observed distance moduli, normalized to $H_0^{\text{ref}}$. Temporarily restoring $c$ for clarity, and assuming the validity of the distance-duality relation, we then construct the dimensionless luminosity and transverse comoving distances $\widetilde D_L=10^{(\mu-25)/5}H_0^{\text{fid}}/c$ and $\widetilde D_M=\widetilde D_L/(1+z)$. We note that this operation explicitly cancels the dependence on $H_0^{\text{fid}}$, confirming that the introduction thereof is required solely to place the different datasets on a common but otherwise completely arbitrary dimensionless distance scale. We also stress that the transformation applied to the distance moduli of all three SNeIa samples are propagated to the corresponding covariance matrices.

A similar procedure needs to be carried out with BAO measurements, in order to remove information on their absolute normalization and keep only shape information (see also Ref.~\cite{Zhang:2025bmk}). To do so, we multiply $D_M/r_d$ and $D_H/r_d$ by $\alpha \equiv H_0r_d/c$, parametrized in terms of $r_dh$. The crucial point is that we do \textit{not} fix this amplitude either through an early-time calibration for $r_d$, or an external prior on $H_0$. We instead treat $r_dh$ as a nuisance parameter to be inferred from data and marginalized over alongside the GP hyperparameters, as part of the GP regression analysis described below. Any redshift-independent rescaling of our BAO distances is absorbed by $r_dh$ and does not, by construction, affect the shape of the reconstructed dimensionless expansion history. Our procedure effectively allows us to treat BAO measurements as a relative, as opposed to absolute, distance indicator. In this sense, our BAO measurements contain purely shape information. We stress once more that our ``rescaled'' BAO measurements retain information on the consistency between $D_H(z)$ and $D_M(z)$, since $\widetilde D_H(z)=\widetilde D_M'(z)$.

With SNeIa and BAO measurements now placed on a common scale, we perform a joint GP regression of $\widetilde D_M(z)$ and its derivative $\widetilde D_M'(z)=\widetilde D_H(z)$, using a modified version of \texttt{GaPP}~\cite{Seikel:2012uu,Bengaly:2020vly}. SNeIa and transverse BAO measurements enter in the reconstruction of the former, and line-of-sight BAO measurements in the latter, although the full cross-covariance between transverse and radial measurements is adopted. We also impose the physically motivated boundary conditions $\widetilde D_M(0)=0$ and $\widetilde D_M'(0)=1$. We adopt a squared-exponential kernel with amplitude $\sigma_f$ and correlation length $\ell$. While we set the GP prior mean function to zero, this should not be confused with the posterior mean reconstruction, which is non-zero once we condition the GP on the data.

We then perform a joint inference of the GP hyperparameters and the BAO normalization. We use Markov Chain Monte Carlo (MCMC) methods, through the \texttt{MontePython} cosmological MCMC sampler~\cite{Audren:2012wb,Brinckmann:2018cvx}, to sample the joint $\sigma_f$-$\ell$-$r_dh$ posterior distribution (or $\sigma_f$-$\ell$ for the three reconstructions which only use SNeIa data). For each sample resulting from the posterior, we condition the GP on the data to draw realizations of $\widetilde D_M'(z)$. We stress that this procedure samples both levels of our hierarchical inference, i.e.\ the uncertainty in the GP hyperparameters and BAO normalization via the MCMC chains, and the (correlated) uncertainty in the reconstructed function via our draws of $\widetilde D_M'(z)$, in a way which preserves the correlations between different redshifts. For each realization we then derive $E(z)=1/\widetilde D_M'(z)$, and compute $I_<$ by integrating $1/E(z)$ up to the appropriate $z_{\max}$. Further technical details on our GP regression methodology are provided in Appendix~\ref{app:gpmethodology}. As a consistency check, we also adopt an alternative node-based reconstruction methodology, which still makes use of GPs, although not in the standard regression sense. The results, presented in Appendix~\ref{app:alternativegpmethodology}, are in excellent agreement with those obtained from our baseline reconstruction methodology and discussed in the main text.

\section{Results}
\label{sec:results}

\begin{table*}[!htbp]
\footnotesize
\resizebox{0.9\textwidth}{!}{
\begin{tabular}{|c?c|c||c|c||c|}
\hline
\multirow{2}{*}{\textbf{Dataset}} & \multirow{2}{*}{$z_{\max}$} & \multirow{2}{*}{$f_<$} & \multicolumn{2}{c||}{$\delta I_</I_<^{\text{fid}}$ ($\%$)} & $(\delta H_0/H_0)_<$ \\
\cline{4-6}
& & & Credible interval & Upper $68\%$ $(95\%)$ & Upper $68\%$ $(95\%)$ \\
\hline\hline
\textit{PantheonPlus} & $2.26$ & $0.41$ & $1.0^{+5.0(+11.0)}_{-6.0(-10.0)}$ & $6.0 \quad (12.0)\%$ & $2.5 \quad (4.9)\%$ \\ \hline
\textit{DESY5Dvk} & $1.13$ & $0.27$ & $-4.0^{+2.1(+3.4)}_{-1.6(-4.0)}$ & $-1.9 \quad (-0.6)\%$ & $-0.51 \quad (-0.16)\%$ \\ \hline
\textit{Union3} & $2.26$ & $0.41$ & $-2.8^{+4.2(+11.0)}_{-6.5(-9.9)}$ & $1.4 \quad (8.2)\%$ & $0.57 \quad (3.4)\%$ \\ \hline \hline
\textit{PantheonPlus}+\textit{DESI} & $2.33$ & $0.42$ & $-0.22^{+0.86(+1.7)}_{-0.86(-1.7)}$ & $0.64 \quad (1.5)\%$ & $0.27 \quad (0.63)\%$ \\ \hline
\textit{DESY5Dvk}+\textit{DESI} & $2.33$ & $0.42$ & $-0.61^{+0.76(+1.5)}_{-0.76(-1.5)}$ & $0.15 \quad (0.9)\%$ & $0.06 \quad (0.38)\%$ \\ \hline
\textit{Union3}+\textit{DESI} & $2.33$ & $0.42$ & $-1.0^{+1.5(+2.9)}_{-1.5(-3.1)}$ & $0.5 \quad (1.9)\%$ & $0.21 \quad (0.80)\%$ \\ \hline \hline
\textit{DESI} & $2.33$ & $0.42$ & $-0.4^{+2.8(+5.0)}_{-1.8(-5.8)}$ & $2.4 \quad (4.6)\%$ & $1.0 \quad (1.9)\%$ \\ \hline
\end{tabular}}
\caption{Constraints on $\delta I_</I_<^{\text{fid}}$, percentage relative deviation in the dimensionless distance integral evaluated up to $z_{\max}$ [see Eq.~(\ref{eq:ilig})], obtained from different combinations of our SNeIa and BAO datasets. We report both $68\%$ and $95\%$ credible intervals (fourth column), as well as the corresponding $68\%$ and $95\%$ upper limits (fifth column). For each dataset combination, we also report both $z_{\max}$ (second column) and the weight $f_<$ (third column), as defined in Eq.~(\ref{eq:weights}). The rightmost column translates the upper limits for $\delta I_</I_<$ into upper limits on the directly constrained low-redshift contribution to the percentage relative deviation in $H_0$, $(\delta H_0/H_0)_< \equiv f_<\delta I_</I_<^{\text{fid}}$, discussed in Sec.~\ref{sec:discussion}.}
\label{tab:deltailpercentage}
\end{table*}

\begin{figure*}[!htbp]
\centering
\includegraphics[width=0.31\linewidth]{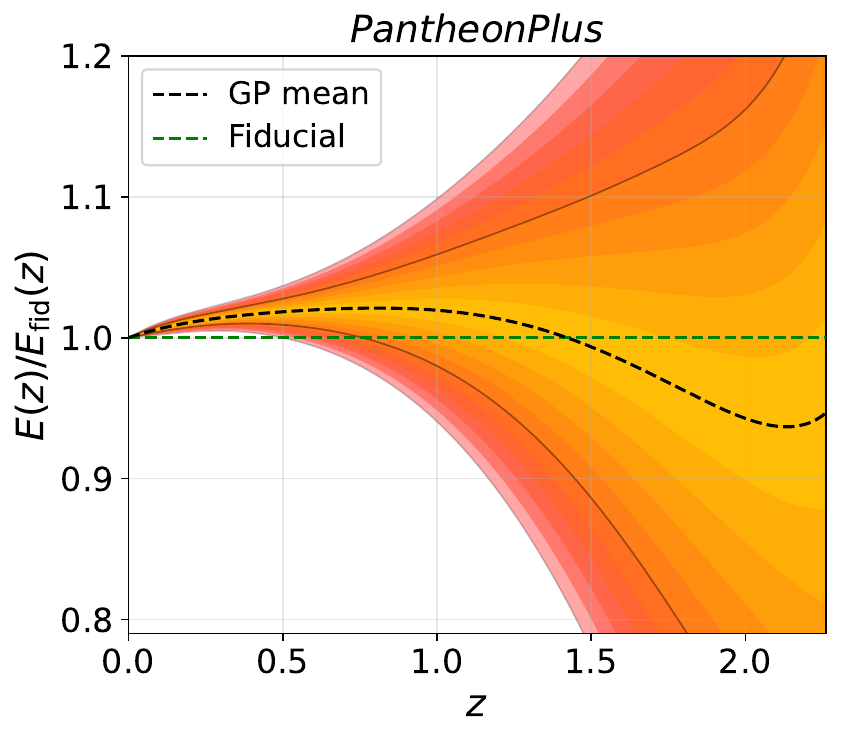} \quad \includegraphics[width=0.31\linewidth]{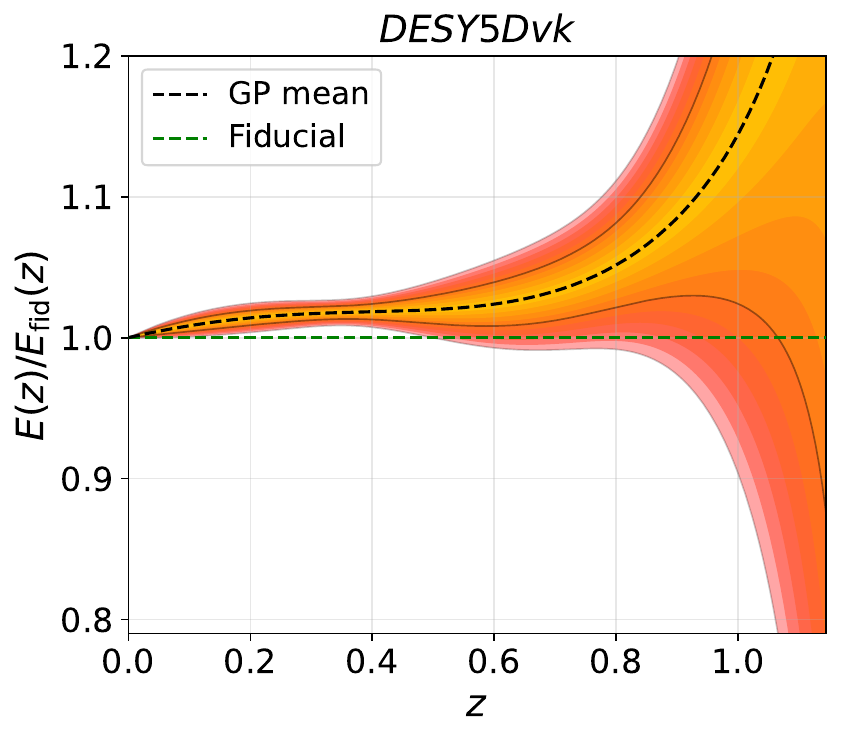} \quad \includegraphics[width=0.31\linewidth]{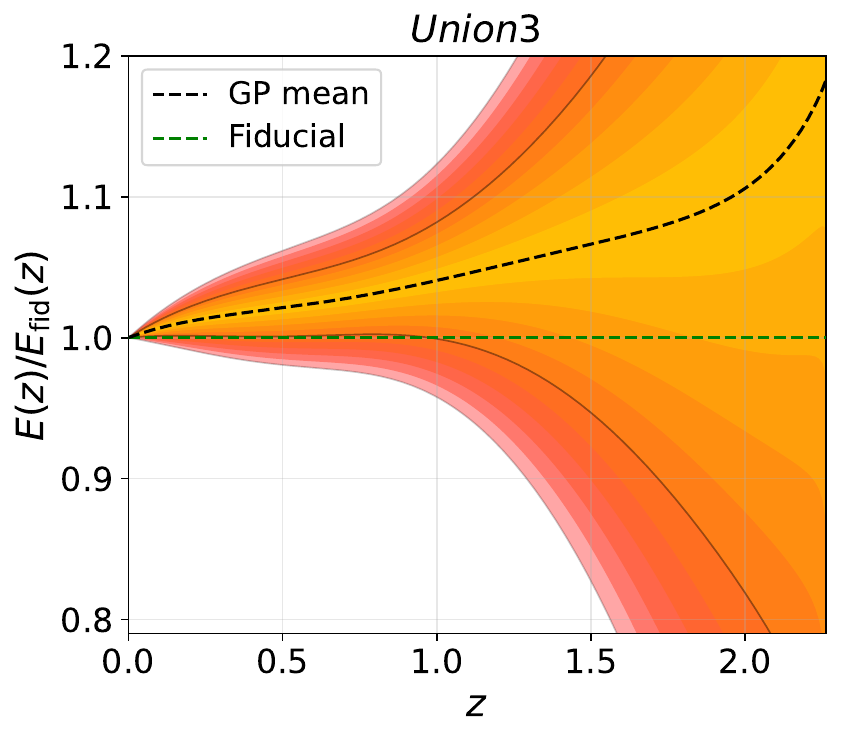} \\
\includegraphics[width=0.31\linewidth]{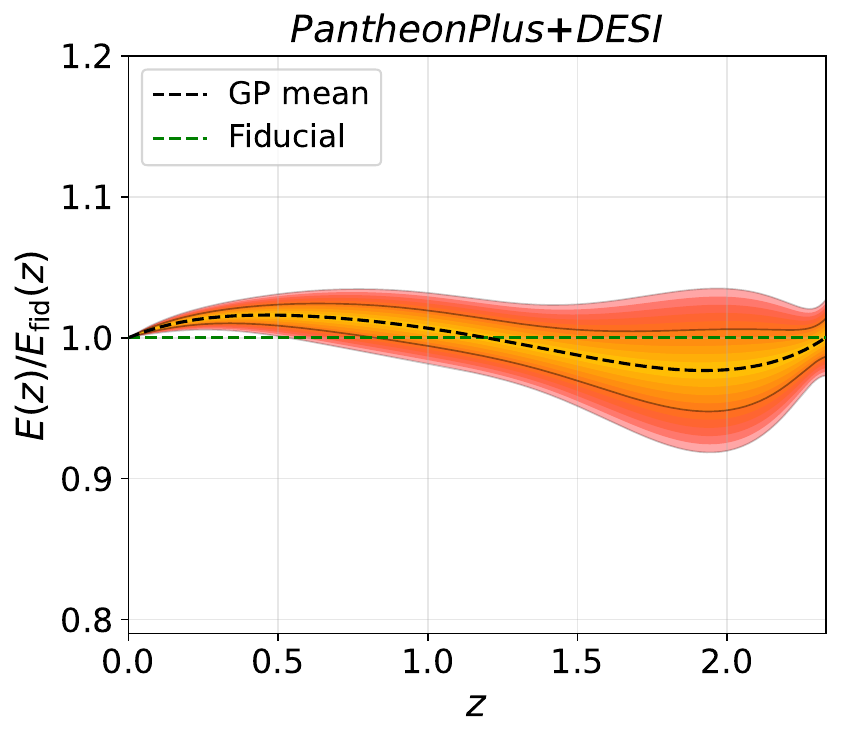} \quad \includegraphics[width=0.31\linewidth]{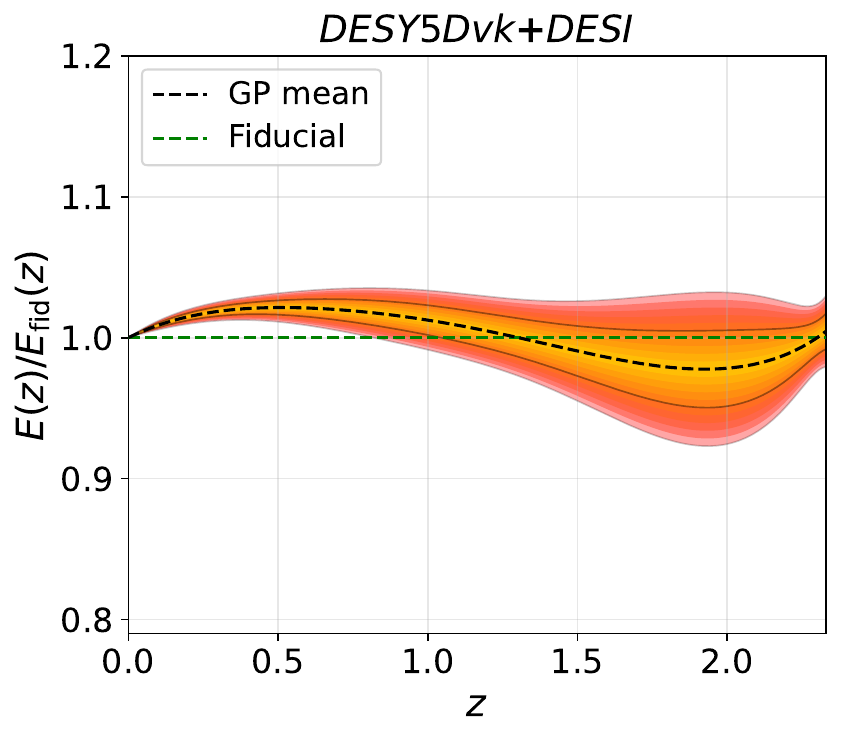} \quad \includegraphics[width=0.31\linewidth]{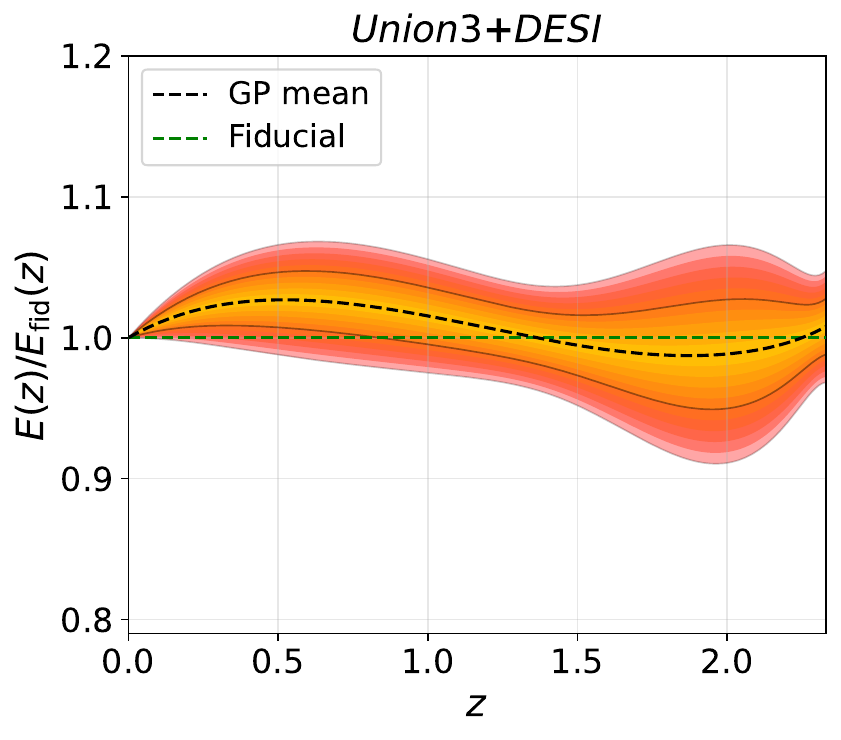}
\caption{Posterior reconstructions of the dimensionless expansion rate relative to the fiducial $\Lambda$CDM model, $E(z)/E_{\text{fid}}(z)$, obtained through Gaussian Process regression. The black dashed curves correspond to the posterior mean reconstruction, whereas the inner and outer pairs of gray thin curves delimit pointwise $68\%$ and $95\%$ credible intervals, and the green horizontal line corresponds to the fiducial model, i.e.\ $E(z)/E_{\text{fid}}(z)=1$. The results are obtained in light of the three SNeIa-only dataset combinations (three upper sub-panels), and the three SNeIa+BAO dataset combinations (three lower sub-panels). From the latter, which we deem as most trustworthy and treat as our baseline dataset combinations, we see that $E(z)$ is constrained to deviate by no more than $\approx 5\%$ from the fiducial model.}
\label{fig:reconstructions}
\end{figure*}

\begin{figure*}[htpb!]
\centering
\includegraphics[width=0.31\linewidth]{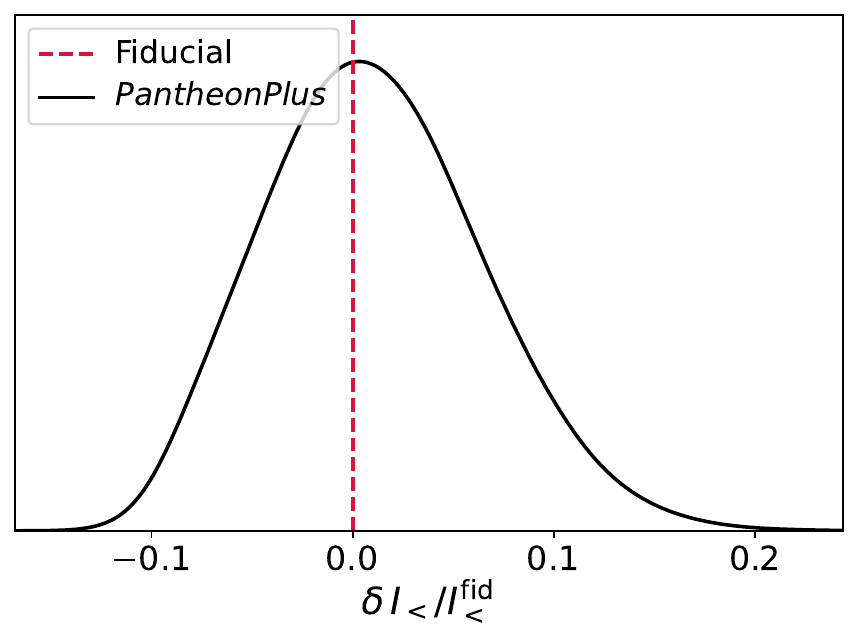} \quad \includegraphics[width=0.31\linewidth]{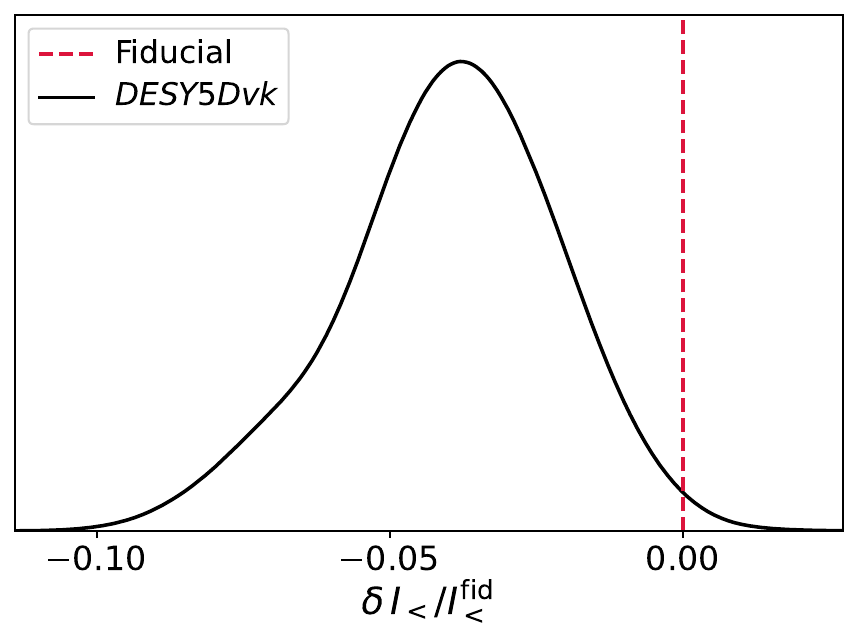} \quad \includegraphics[width=0.31\linewidth]{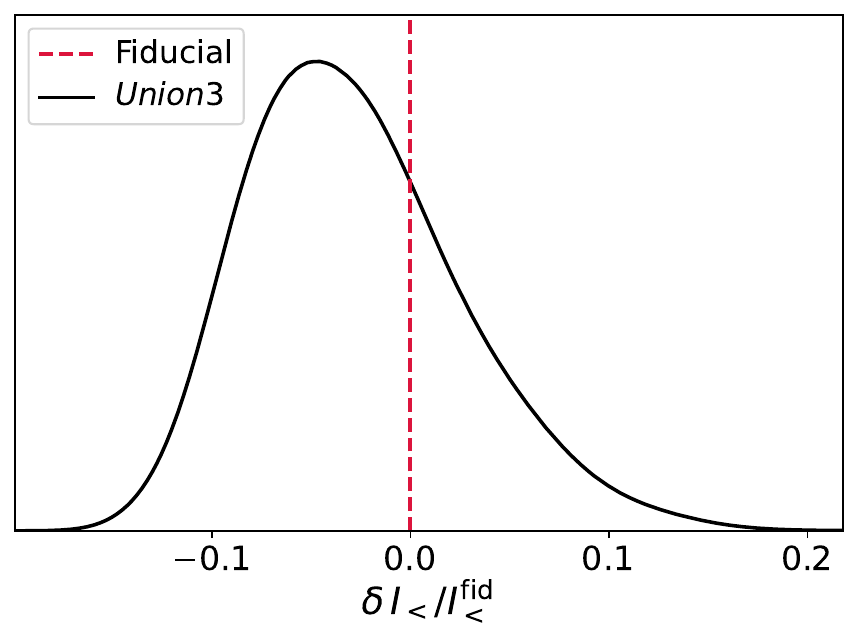} \\
\includegraphics[width=0.31\linewidth]{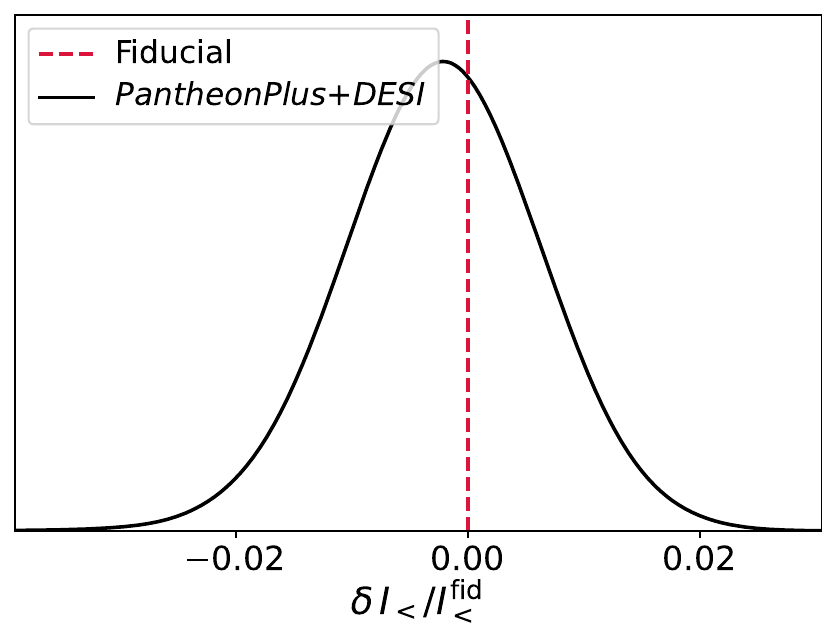} \quad \includegraphics[width=0.31\linewidth]{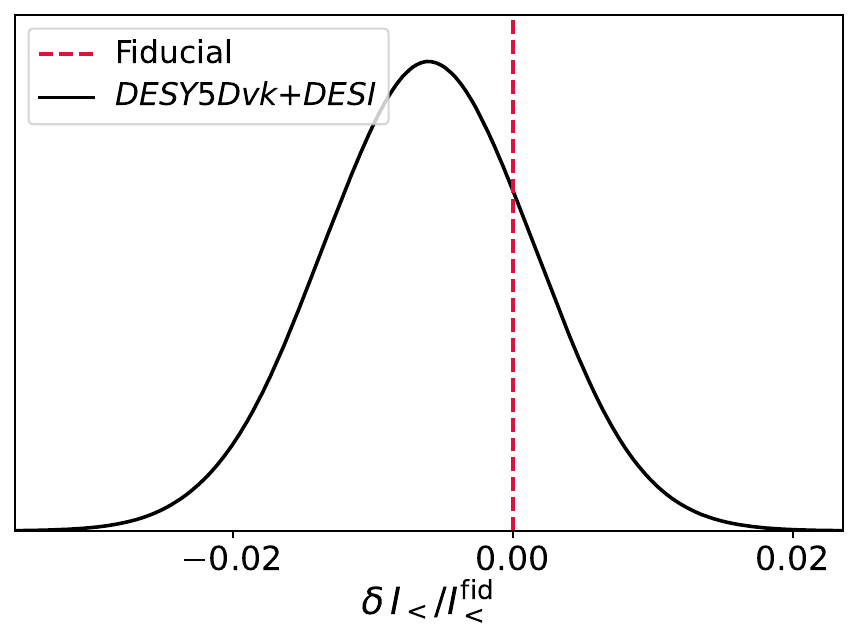} \quad \includegraphics[width=0.31\linewidth]{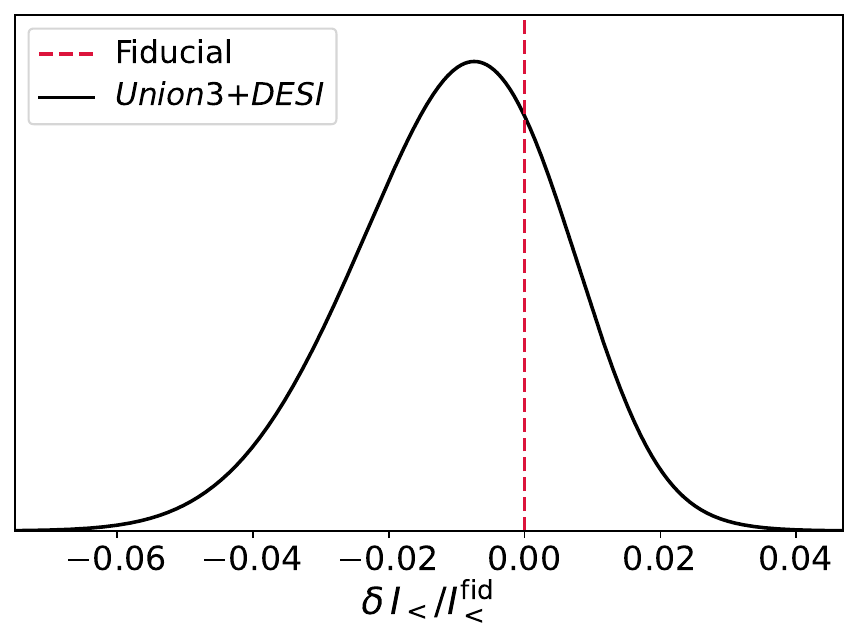}
\caption{1D marginalized posterior distributions for $\delta I_</I_<^{\text{fid}}$, the relative shift in the dimensionless distance integral allowed for $z<z_{\max}$, obtained in light of the three SNeIa-only dataset combinations (three upper sub-panels), and the three SNeIa+BAO dataset combinations (three lower sub-panels). Summary statistics for these posterior distributions are reported in Tab.~\ref{tab:deltailpercentage}. Note that $z_{\max}$ depends on the adopted dataset combination (see the second column in Tab.~\ref{tab:deltailpercentage}).}
\label{fig:posteriorsdeltail}
\end{figure*}

We now discuss the results obtained using the methodology and datasets presented in Sec.~\ref{sec:datasets}. Our reconstructed dimensionless expansion rates normalized to the fiducial $\Lambda$CDM model, i.e.\ $E(z)/E_{\text{fid}}(z)$, are presented in Fig.~\ref{fig:reconstructions}, with each sub-panel referring to one of the six dataset combinations discussed in Sec.~\ref{sec:datasets}. The corresponding posterior distributions for $\delta I_</I_<^{\text{fid}}$ are shown in Fig.~\ref{fig:posteriorsdeltail}, whereas constraints on $\delta I_</I_<^{\text{fid}}$ (both credible intervals and upper $68\%$ and $95\%$ limits) are reported in Tab.~\ref{tab:deltailpercentage}, alongside the relevant values of $z_{\max}$ and weight $f_<$ discussed earlier [see Eq.~(\ref{eq:weights})]. For later reference, the rightmost column translates the upper limits for $\delta I_</I_<$ into upper limits on the directly constrained low-redshift contribution $(\delta H_0/H_0)_< \equiv f_<\delta I_</I_<^{\text{fid}}$: we will not discuss the results reported in this column here, but will return to their implications in Sec.~\ref{sec:discussion}.

We first inspect the reconstructions of $E(z)/E_{\text{fid}}(z)$ shown in Fig.~\ref{fig:reconstructions}. For each panel, the black dashed curve corresponds to the posterior mean reconstruction, whereas the inner and outer pairs of gray thin curves delimit pointwise $68\%$ and $95\%$ credible intervals (see Appendix~\ref{app:gpmethodology} for further technical details). We recall that all reconstructions satisfy $E(0)/E_{\text{fid}}(0)=1$ by construction. At the qualitative level we observe a mild rise above $E_{\text{fid}}(z)$ at $z \lesssim 0.5$, followed by a dip below $E_{\text{fid}}(z)$ at $z \gtrsim 1.5$, in all but the \textit{DESY5Dvk}- and \textit{Union3}-only reconstructions, which instead rise continuously towards higher redshifts. These qualitative features, which become somewhat more pronounced when \textit{DESI} BAO data is included, are in very good agreement with similar features identified in other independent nonparametric reconstructions of $E(z)$~\cite{Jiang:2024xnu,Kessler:2026dbi,GuptaChoudhury:2026gsl}. At high redshift, the SNeIa-only reconstructions show significant freedom, with the corresponding pointwise credible intervals widening substantially.

Given that BAO measurements are sparse in redshift compared to SNeIa, the reduction in reconstruction uncertainty once BAO data are added (compare the three lower versus three upper sub-panels in Fig.~\ref{fig:reconstructions}) may appear somewhat surprising. To see why this occurs, let us recall that SNeIa only constrain $\widetilde D_M(z)$, i.e. the integral of $1/E(z)$. Recovering $E(z)$ therefore requires differentiating the reconstructed distance, which inevitably amplifies the reconstruction uncertainties, especially towards $z_{\max}$. On the other hand, line-of-sight BAO measurements directly constrain $E(z)$, once we marginalize over the normalization. Combining line-of-sight and transverse BAO measurements therefore constraints both distances and their derivatives, which is precisely what is required to ``stabilize'' the reconstruction of $E(z)$. Another point worth noting is that the uncertainties in SNeIa measurements increase significantly for $z \gtrsim 1$ (see for instance the binned measurements in Fig.~1 of Ref.~\cite{Poulin:2024ken}, where this trend is extremely clear). Therefore, measurements, which extend up to $z=2.33$, add significant constraining power precisely where SNeIa measurements are weakest. Combining these considerations, it becomes very clear why BAO measurements significantly help in sharpening the reconstruction of the dimensionless expansion history, despite their sparseness in redshift. For these reasons, we deem the results obtained from our three SNeIa+BAO dataset combinations (i.e.\ \textit{PantheonPlus}+\textit{DESI}, \textit{DESY5Dvk}+\textit{DESI}, and \textit{Union3}+\textit{DESI}) as being our most trustworthy ones, and will treat them as our baseline results, while also discussing the SNeIa-only results for completeness.

We find all our baseline SNeIa+BAO reconstructions (three lower sub-panels of Fig.~\ref{fig:reconstructions}) to be in excellent agreement between each other, whereas BAO data is especially helpful in reducing the reconstruction uncertainties at $z \gtrsim 1$. All three reconstructions display the feature in $E(z)$ discussed earlier which, within a spatially flat matter-plus-dark energy interpretation, would correspond to an effective dark energy equation of state crossing the phantom divide, going from $w>-1$ at low redshifts to $w<-1$ at higher redshifts, as widely discussed in the recent literature (see e.g.\ Refs.~\cite{Cortes:2024lgw,Colgain:2024xqj,Carloni:2024zpl,Giare:2024gpk,Jiang:2024viw,Giare:2024oil,Giare:2025pzu,Colgain:2025nzf,DESI:2025wyn,RoyChoudhury:2025dhe,Afroz:2025iwo,Ozulker:2025ehg,Li:2025dwz,Chaudhary:2025pcc,Chaudhary:2025vzy,Fazzari:2025lzd,Chaudhary:2025uzr,RoyChoudhury:2025iis,Li:2025muv,Li:2025vuh,Capozziello:2025qmh,SolaPeracaula:2026lyk}). The precise amplitude of these features depends on the SNeIa compilation. For instance, the \textit{DESY5Dvk}+\textit{DESI} reconstruction displays the most statistically significant deviation from $E_{\text{fid}}(z)$ at low redshifts, whereas the \textit{Union3}+\textit{DESI} reconstruction shows a more pronounced low-redshift rise, but with correspondingly larger credible intervals. These results are in agreement with independent findings in the literature. The consistency we observe across the different reconstructions is important, as it indicates that the main features we observe are not an artifact of the choice of SNeIa dataset. Broadly speaking, our SNeIa+BAO reconstructions indicate that the shape of the dimensionless expansion history can deviate, at best, by no more than $\approx 5$-$7\%$ from the shape of the fiducial model, with the largest deviations being allowed at redshifts $1.5 \lesssim z \lesssim 2$. At redshifts $z \lesssim 1.5$, relative deviations in the shape of the expansion history are instead tightly constrained at the $\lesssim 5\%$ level.

From these reconstructions, we now derive posterior distributions for $\delta I_</I_<^{\text{fid}}$. Given that $I_<$ is obtained by integrating $1/E(z)$, in principle deviations at different redshifts could either reinforce or compensate each other. In particular, the low-redshift rise in $E(z)$ will contribute negatively to $I_<$, and conversely for the dip at higher redshifts. To obtain the posterior for $\delta I_</I_<^{\text{fid}}$, we compute the $I_<$ integral separately for each realization, and the resulting ensembles of $I_</I_<^{\text{fid}}$ values define the posterior distributions shown in Fig.~\ref{fig:posteriorsdeltail}. We stress that this procedure automatically accounts for correlations across different redshifts in the reconstruction. It is important to note that the credible envelopes shown in Fig.~\ref{fig:reconstructions} are \textit{pointwise}, and do not in general correspond to individual reconstructed histories. In other words, their values at different redshifts in general originate from different realizations of $E(z)/E_{\text{fid}}(z)$. Therefore, one \textit{cannot} obtain credible intervals for $\delta I_</I_<^{\text{fid}}$ by simply integrating the corresponding pointwise envelopes. For instance, the integral of the inverse of the $68\%$ upper pointwise envelope for $E(z)$ will not map onto the $68\%$ lower limit on $I_<$, and similarly for the corresponding lower pointwise envelope and $68\%$ upper limit. We will return to this subtle point and its implications in Sec.~\ref{sec:discussion}.

In Tab.~\ref{tab:deltailpercentage}, we report both $68\%$ and $95\%$ credible intervals for $\delta I_</I_<^{\text{fid}}$, as well as the corresponding $68\%$ and $95\%$ upper limits. We focus only on upper rather than lower limits, as these are the ones which are of direct interest to this work, given that they can be mapped [through Eqs.~(\ref{eq:deltah0},\ref{eq:deltaildeltaig})] onto the maximum relative increase in $H_0$ which is allowed by direct, late-time constraints on the shape of the expansion history. Our SNeIa-only results are in principle rather broad, although they depend significantly on the specific sample adopted. For instance, the \textit{PantheonPlus} central value is $\delta I_</I_<^{\text{fid}}=1\%$, but the comparatively large uncertainties in the reconstruction (see the upper left sub-panel of Fig.~\ref{fig:reconstructions}) map onto upper $68\%$ and $95\%$ limits of $\delta I_</I_<^{\text{fid}}=6\%$ and $12\%$ respectively. Conversely, the central values of $\delta I_</I_<^{\text{fid}}$ obtained from the \textit{DESY5Dvk} and \textit{Union3} reconstructions are negative. This is consistent with the shape of the distance-redshift relation preferred by these two samples. In particular, both datasets are known to mildly prefer a thawing-like dark energy evolution, as well as larger values of $\Omega_m$ compared to our fiducial value of $0.31$, which is part of a more general mild disagreement between late-time probes for what concerns the value of $\Omega_m$~\cite{Baryakhtar:2024rky,Pedrotti:2024kpn,Colgain:2024mtg,Weiner:2026sfm,Shlivko:2026jxa,Schoneberg:2026buf}. At any rate, both trends lead to a larger expansion rate over the relevant redshift range (as is evident from the upper middle and right sub-panels in Fig.~\ref{fig:reconstructions}), and translates into a negative contribution to $\delta I_</I_<^{\text{fid}}$. As stressed earlier, our baseline and most trustworthy results are those obtained from the three SNeIa+BAO dataset combinations. In this case, what we see is that a much more coherent picture emerges, from which we can draw three general conclusions:
\begin{itemize}
\item regardless of the SNeIa sample being adopted, all SNeIa+BAO constraints tighten substantially, by more than half an order of magnitude or more, compared to their SNeIa-only counterparts;
\item all three central values of $\delta I_</I_<^{\text{fid}}$ are negative;
\item all three constraints are broadly in agreement with each other, and point towards rather tight $68\%$ ($95\%$) upper limits on $\delta I_</I_<^{\text{fid}}$ of $\approx 0.5\%$ ($\approx 1.5\%$).
\end{itemize}
For what concerns the last point, the $68\%$ ($95\%$) upper limits on $\delta I_</I_<^{\text{fid}}$ we find from \textit{PantheonPlus}+\textit{DESI}, \textit{DESY5Dvk}+\textit{DESI}, and \textit{Union3}+\textit{DESI} are $0.64\%$ ($1.5\%$), $0.15\%$ ($0.9\%$), and $0.5\%$ ($1.9\%$) respectively.~\footnote{{We refer the reader to Appendix~\ref{app:omegamdependence} for a dependency test against variations in the fiducial value of $\Omega_{\rm m}$.}} For completeness, in Tab.~\ref{tab:deltailpercentage} we also report the values obtained from a \textit{DESI}-only reconstruction: although this is not shown in Fig.~\ref{fig:reconstructions}, the features are very similar to those observed in the SNeIa+BAO reconstructions, only with uncertainties roughly twice as large. For what concerns $\delta I_</I_<^{\text{fid}}$, the results are somewhat intermediate between the SNeIa-only and SNeIa+BAO ones, with $\delta I_</I_<^{\text{fid}}<2.4\%$ ($4.6\%$) at the $68\%$ ($95\%$) level.

All in all, our results indicate that shape constraints on the post-recombination expansion history are extremely tight, and overall in very good agreement with $\delta I_</I_<^{\text{fid}}=0$. Once complementary shape information from SNeIa and BAO is combined, the allowed positive variation in the (unweighted) part of the dimensionless distance integral which is directly constrained by low-redshift data is restricted to the percent-level at best. This is the main quantitative result of the first part of our analysis: \textit{over the redshift range directly covered by late-time observations, there is a percent-level shape wall for $I_<$, whose exact value depends only marginally on the adopted SNeIa sample}. We stress that this does not yet provide a constraint on the full allowed shift in $H_0$, since we still have to introduce the $f_<$ weight, as well as the high-redshift contribution from $f_>I_>$, which is not directly constrained by data. However, already at this stage we can expect that the resulting constraints on $\delta H_0/H_0^{\text{fid}}$ will be particularly tight. These statements will be made more quantitative, and critically discussed, in Sec.~\ref{sec:discussion}.

\section{Discussion}
\label{sec:discussion}

We now critically discuss the implications of our earlier results for the Hubble tension. From Eqs~(\ref{eq:deltah0},\ref{eq:deltaildeltaig}), we see that our upper limits on $\delta I_</I_<^{\text{fid}}$ can be translated into the contribution to $\delta H_0/H_0^{\text{fid}}$ which is directly constrained by late-time observations. This contribution, which we refer to as $(\delta H_0/H_0)_<$, is given as follows:
\begin{equation}
\left ( \frac{\delta H_0}{H_0} \right ) _< \equiv f_< \frac{\delta I_<}{I_<^{\text{fid}}}\,.
\label{eq:deltah0h0l}
\end{equation}
We report $68\%$ and $95\%$ upper limits on $(\delta H_0/H_0)_<$ in the rightmost column of Tab.~\ref{tab:deltailpercentage} (note that $f_{<}$ is given, for each dataset combination, in the third column). Because of the low-redshift weight $f_{<}$, which is $f_{<} \simeq 0.42$ for all SNeIa+BAO dataset combinations, the already very tight constraints on $\delta I_</I_<^{\text{fid}}$ discussed earlier map onto even tighter constraints on $(\delta H_0/H_0)_<$. Even in the least constraining case among the three, i.e.\ the \textit{Union3}+\textit{DESI} one, the $95\%$ upper limit on $(\delta H_0/H_0)_<$ is a remarkably tight $0.8\%$, an order of magnitude short of the $\approx 9\%$ shift required to fully solve the Hubble tension. The most constraining $95\%$ upper limit of $0.38\%$ is instead inferred from the \textit{DESY5Dvk}+\textit{DESI} dataset combination, and is caused by the low-redshift feature extensively discussed in Sec.~\ref{sec:results}, and evident in the lower right sub-panel of Fig.~\ref{fig:reconstructions}. In short, we conclude that the shape of the late-time expansion history directly constrained by low-redshift observations leaves very little room for late-time new physics to increase $H_0$ via modifications to $I_<$. Over the redshift range directly covered by late-time observations, the shape wall for $H_0$ is therefore at the sub-percent level.

One may legitimately wonder why allowed variations of order $\approx 5\%$ in $E(z)$, as observed in the reconstruction upper and lower pointwise envelopes within the three lower sub-panels in Fig.~\ref{fig:reconstructions}, translate into sub-percent constraints on $(\delta H_0/H_0)_<$. The reason, as we somewhat alluded to earlier, is that the displayed envelopes are \textit{pointwise}. In particular, the lower $68\%$ and $95\%$ envelopes are obtained by selecting the corresponding quantile \textit{at each redshift}. Therefore, the quantile values at different redshifts need not belong to the same reconstructed GP realization: in fact, in general, they will not. Stated differently, the region bounded by the pointwise $68\%$ envelope does not correspond to a simultaneous $68\%$ credible region for the \textit{full} reconstructed function. The fraction of GP realizations which fall within this region at all redshifts will, in general, be well below $68\%$. In other words, considering an expansion history close to the lower $68\%$ pointwise envelope across the entire redshift range amounts to considering one which has negligible posterior probability. Integrating the inverse of this dimensionless expansion history would indeed lead to a comparatively large value of $\delta I_</I_<^{\text{fid}} \approx {\cal O}(10\%)$, but such a value would lie well within the upper tails of the posterior distributions of $\delta I_</I_<^{\text{fid}}$ shown in Fig.~\ref{fig:posteriorsdeltail}. On the other hand, our approach towards constructing these posteriors, which integrates each reconstructed GP realization separately, keeps the information about the correlations between different redshifts, and therefore drastically reduces the apparent ``pointwise freedom'' one would otherwise na\"{i}vely infer from the reconstructions. In passing, we note that the integral of the inverse of the posterior mean reconstruction will, in general, provide a reasonable approximation to the central value of the $I_<$ posterior:~\footnote{The difference between the two can be shown to be second-order in the reconstruction uncertainty.} however, the two need not coincide, since the mapping between $E(z)$ and $I_<$ is non-linear. In short, there is no contradiction between the allowed variations of order $\approx 5\%$ in $E(z)$, and the sub-percent, extremely tight constraints on $(\delta H_0/H_0)_<$.

The main remaining caveat is that our reconstruction can only directly constrain $I_<$, rather than the full integral $I=I_<+I_>$ which controls the acoustic angular scale. Since $f_< \simeq 0.42$ for our baseline SNeIa+BAO dataset combinations we are, loosely speaking, only constraining $42\%$ of the allowed increase in $H_0$. More than half of the dimensionless distance integral is not directly constrained. In principle, one could expect that a sufficiently large and positive value of $\delta I_>/I_>^{\text{fid}}$ could compensate the tight constraints on $\delta I_</I_<^{\text{fid}}$ from the shape wall discussed earlier. This could lead to a larger shift in the full integral, and therefore in $H_0$. One could be tempted to try and reconstruct $E(z)$ above $z_{\max}$. We caution against doing so, as it would amount to a prior-dominated extrapolation, give that no data is present to drive the reconstruction. Nevertheless, it is still possible to provide a reasonable estimate of the residual freedom in $I_>$, by introducing additional physical input.

Our approach is to determine how much $I_>$ is allowed to vary within some of the simplest cosmological models against which late-time observations are usually constrained. We stress that this cannot, by construction, provide a nonparametric estimate of $\delta I_>/I_>^{\text{fid}}$, as it relies on the assumption of a specific model. However, it provides us with a physically motivated estimate of the \textit{typical} size of the variations which are allowed in $I_>$ within smooth and well-studied expansion histories. For concreteness, we consider two of the simplest cosmological models: the standard 6-parameter $\Lambda$CDM model, and its 7-parameter $w$CDM extension, where the (constant) dark energy equation of state $w$ is treated as a free parameter (see Ref.~\cite{Escamilla:2023oce} for a recent comprehensive study on constraints within this model). Within $\Lambda$CDM, the remaining freedom in $I_>$ is entirely captured within the allowed variations in $\Omega_m$, as this is the only parameter which controls the shape of the expansion history. Within $w$CDM, it is instead $\Omega_m$ and $w$ which control the remaining freedom in $I_>$. We constrain these models against the \textit{PantheonPlus}+\textit{DESI} dataset combination by using MCMC methods. Setting $z_{\max}=2.33$, we then treat $\delta I_>/I_>^{\text{fid}}$ as a derived parameter whose distribution we estimate from our MCMC chains, analogously to the procedure we followed for $\delta I_</I_<^{\text{fid}}$.

\begin{table*}[!htbp]
\footnotesize
\resizebox{0.9\textwidth}{!}{
\begin{tabular}{|c?c|c||c|c||c|c|}
\hline
\multirow{2}{*}{\textbf{Treatment of $I_>$}} & \multicolumn{2}{c||}{$\delta I_</I_<^{\text{fid}}$} & \multicolumn{2}{c||}{$\delta I_>/I_>^{\text{fid}}$} & \multicolumn{2}{c|}{$\delta I/I^{\text{fid}}\simeq\delta H_0/H_0$} \\
\cline{2-7} & Upper $68\%$ & Upper $95\%$ & Upper $68\%$ & Upper $95\%$ & Upper $68\%$ & Upper $95\%$ \\
\hline\hline
Conservative & $0.64\%$ & $1.5\%$ & $1.22\%$ & $2.48\%$ & $0.98\%$ & $2.07\%$ \\ \hline
Maximally permissive & $0.64\%$ & $1.5\%$ & $3.88\%$ & $5.24\%$ & $2.52\%$ & $3.67\%$ \\ \hline
\end{tabular}}
\caption{$68\%$ and $95\%$ upper limits on $\delta I_</I_<^{\text{fid}}$ (second and third columns) and $\delta I_>/I_>^{\text{fid}}$ (fourth and fifth columns), and corresponding limits on the percentage relative deviation in $H_0$, $\delta H_0/H_0$ (sixth and seventh columns). The results are presented for two different benchmarks. In the conservative benchmark, we combine the \textit{PantheonPlus}+\textit{DESI} upper limits on $I_<$ with the $\Lambda$CDM-based model-guided estimate of $I_>$ which disregards the central value of the $\delta I_>/I_>^{\text{fid}}$ posterior. In the maximally permissive benchmark, we combine the same \textit{PantheonPlus}+\textit{DESI} upper limits on $I_<$ with the $w$CDM-based model-guided limits on $I_>$, taking into account the non-zero central value of the $\delta I_>/I_>^{\text{fid}}$ posterior. Even in the latter benchmark, the allowed shift in $H_0$ falls well short of the values required to fully solve the tension in the late-time Universe.}
\label{tab:deltaipercentage}
\end{table*}

Within our $\Lambda$CDM and $w$CDM benchmark runs, we find posterior means of $\delta I_>/I_>^{\text{fid}}=1.65\%$ and $2.55\%$ respectively. Our $68\%$ ($95\%$) upper limits on $\delta I_>/I_>^{\text{fid}}$ are instead $2.87\%$ ($4.06\%$) and $3.88\%$ ($5.24\%$) within $\Lambda$CDM and $w$CDM respectively. We see that, at least within these models, the residual freedom in $I_>$ is a fair bit larger than that in $I_<$: this should not come as a surprise, given the absence of direct constraints at $z>z_{\max}$. It can also be instructive to momentarily disregard the central value of the $\delta I_>/I_>^{\text{fid}}$ posteriors, i.e.\ $1.65\%$ and $2.55\%$: this is useful as it separates the residual freedom allowed by the data, i.e.\ the width of the $\delta I_>/I_>^{\text{fid}}$ distribution, from the effect of an offset between the best-fit (parametric) expansion history relative to our chosen fiducial expansion history. In other words, this allows us to estimate how much additional freedom lies in the unreconstructed high-redshift contribution, regardless of the central value thereof. Doing so, we find that the residual freedom in $I_>$ is $1.22\%$ ($2.48\%$) and $1.33\%$ ($2.70\%$) at $68\%$ ($95\%$) level, for the $\Lambda$CDM and $w$CDM models respectively. We stress once more that, unlike our constraints on $I_<$, the values quote above rely on an assumed cosmological model: they should not be considered nonparametric bounds on $\delta I_>/I_>^{\text{fid}}$, but rather reasonable model-guided estimates thereof.

We now turn these figures into an estimate of the \textit{total} allowed variation in the dimensionless distance integral, $\delta I_>/I_>^{\text{fid}}$, using Eq.~(\ref{eq:deltaildeltaig}). In our most conservative setting, we combine the \textit{PantheonPlus}+\textit{DESI} upper limits on $I_<$, i.e.\ $\delta I_</I_<^{\text{fid}} \lesssim 0.64\%$ ($1.5\%$) at $68\%$ ($95\%$), with the $\Lambda$CDM-based model-guided estimate of $I_>$ which disregards the central value of the $\delta I_>/I_>^{\text{fid}}$ posterior, i.e.\ $\delta I_>/I_>^{\text{fid}} \lesssim 1.22\%$ ($2.48\%$). Using the weights $f_< \simeq 0.42$ and $f_> \simeq 0.58$, we find the $68\%$ and $95\%$ upper limits of $\delta I/I^{\text{fid}} \lesssim 0.98\%$ and $2.07\%$ respectively. We have conservatively added the two contributions linearly, assuming that both can simultaneously saturate their respective upper limits. The maximally permissive approach for what concerns the $I_>$ piece of the integral combines the same \textit{PantheonPlus}+\textit{DESI} upper limits on $I_<$ with the $w$CDM-based model-guided limits on $I_>$, taking into account the non-zero central value of the $\delta I_>/I_>^{\text{fid}}$ posterior, i.e.\ $\delta I_>/I_>^{\text{fid}}  \lesssim 3.88\%$ ($5.24\%$). Doing so we find the $68\%$ and $95\%$ upper limits of $\delta I/I^{\text{fid}} \lesssim 2.52\%$ and $3.67\%$ respectively. These figures are summarized in Tab.~\ref{tab:deltaipercentage} (see the second to fifth columns). In a scenario where only the late-time expansion history is modified, and therefore $r_s^{\star}$ is unchanged, Eq.~(\ref{eq:deltah0}) implies these estimates of $\delta I/I^{\text{fid}}$ directly translate into the maximum allowed relative variation in the Hubble constant, $\delta H_0/H_0$. In the most conservative setting, we therefore find $\delta H_0/H_0 \lesssim 2.07\%$ at $95\%$, well below the $\gtrsim 9\%$ increase required to fully solve the tension. Importantly, even within the maximally permissive treatment, our limit of $\delta H_0/H_0 \lesssim 3.67\%$ still falls well short of the values required to fully solve the tension in the late-time Universe. Taking as reference the same H0DN and CMB-based values reported in Sec.~\ref{sec:theoretical}, we see that a $3.67\%$ increase in $H_0$ would still leave a residual Gaussian tension of $\approx 4.3\sigma$ with the H0DN measurement, which is far from acceptable.

\begin{figure*}[!htbp]
\centering
\includegraphics[width=0.31\linewidth]{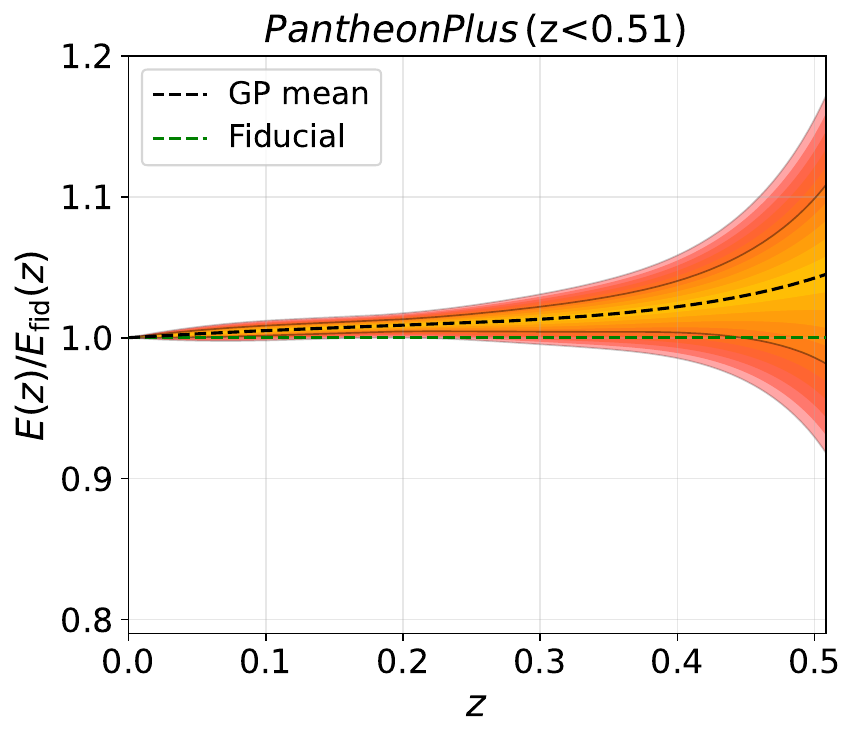} \quad \includegraphics[width=0.31\linewidth]{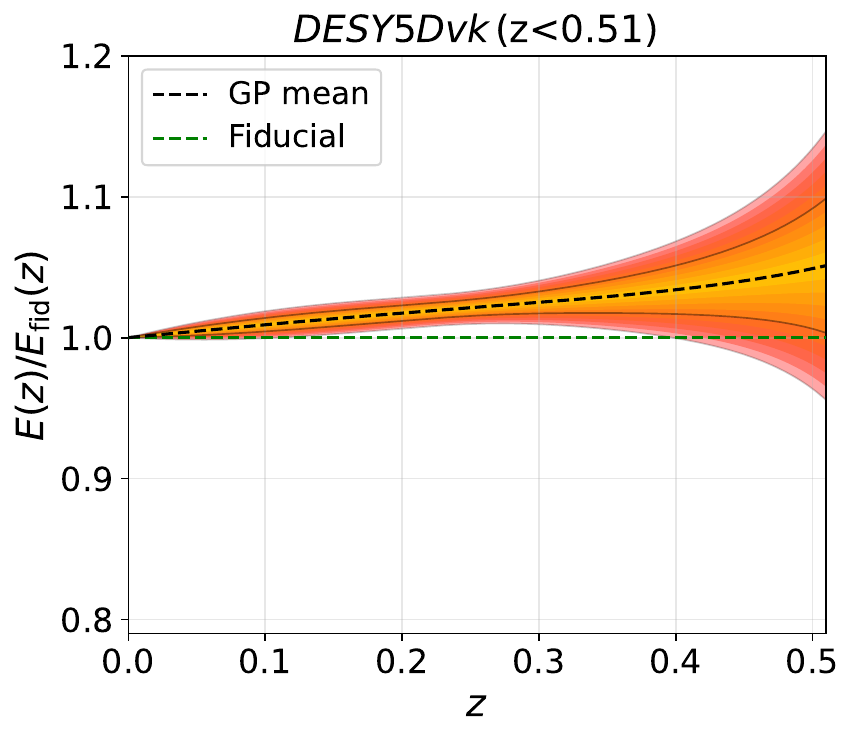} \quad \includegraphics[width=0.31\linewidth]{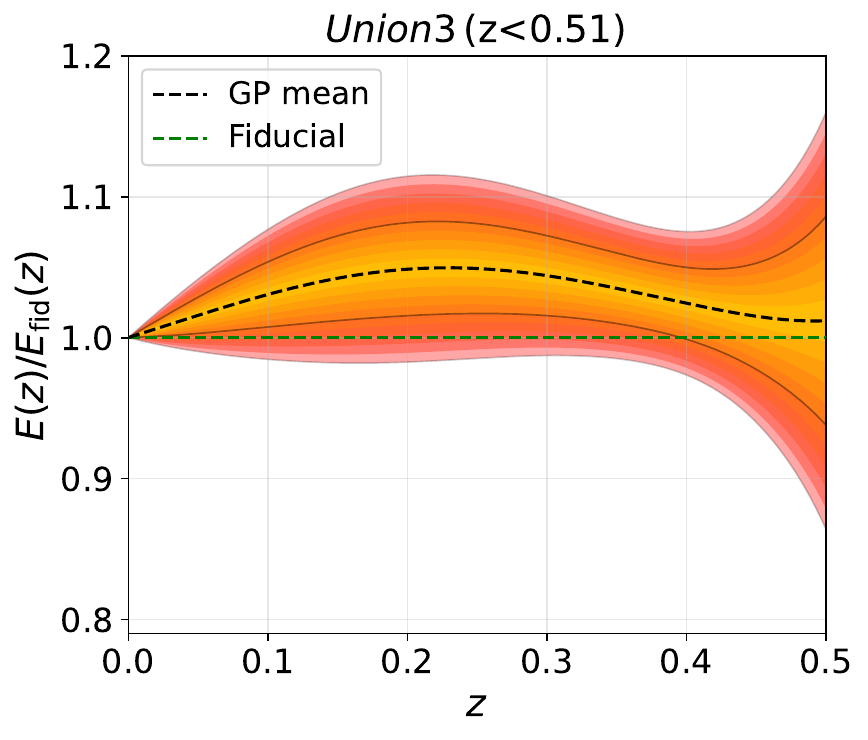}
\caption{As in the three upper sub-panels of Fig.~\ref{fig:reconstructions}, but for the case where only SNeIa for $z<0.51$ are considered.}
\label{fig:reconstructionscut}
\end{figure*}

\begin{figure*}[htpb!]
\centering
\includegraphics[width=0.31\linewidth]{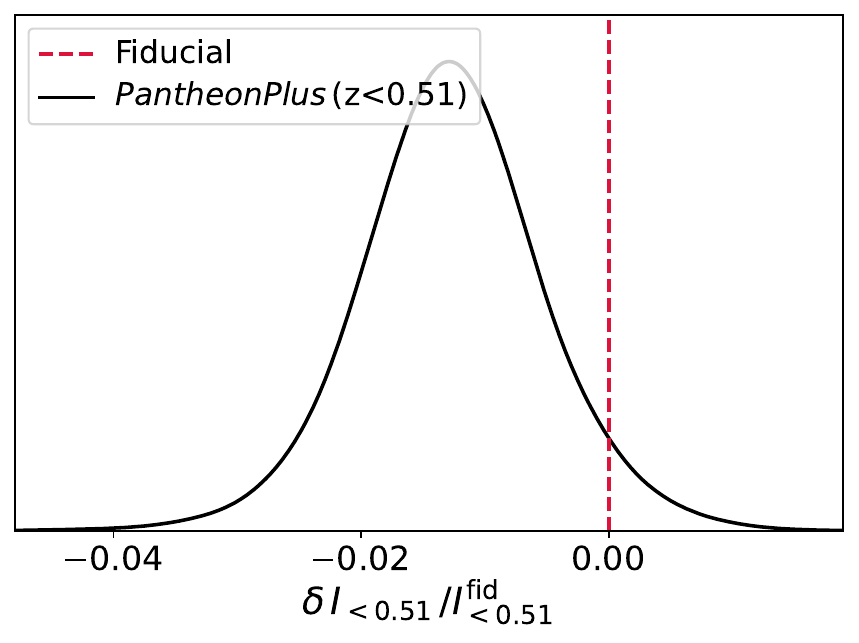} \quad  \includegraphics[width=0.31\linewidth]{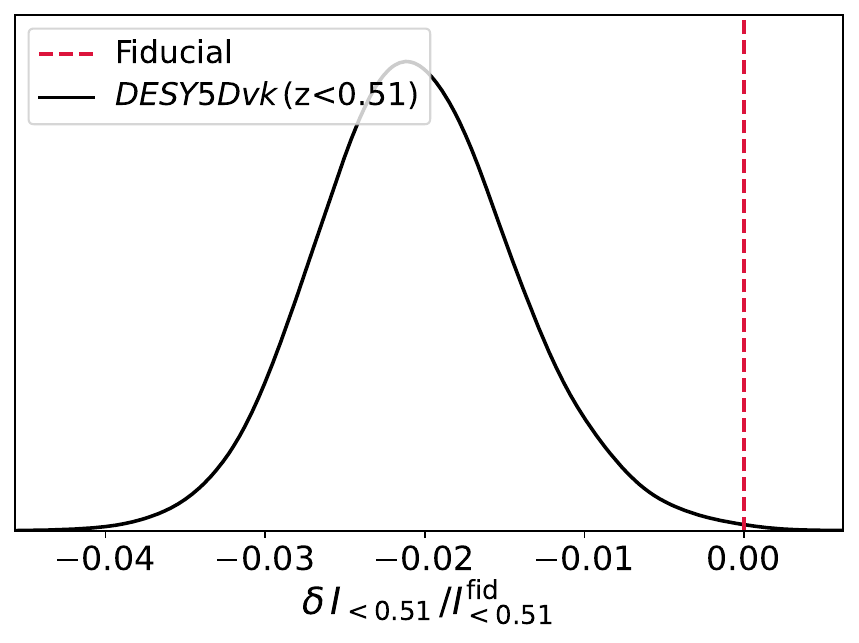} \quad \includegraphics[width=0.31\linewidth]{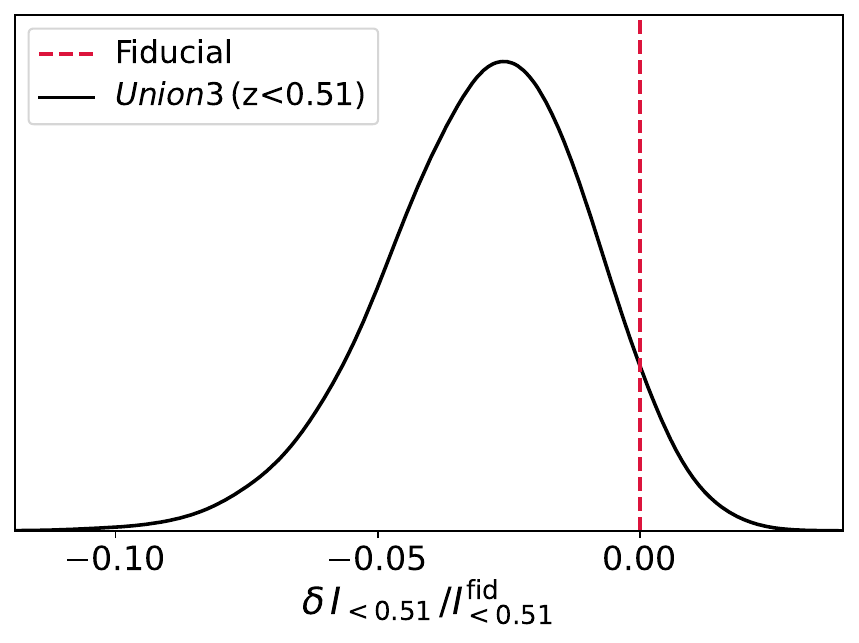}
\caption{As in the three upper sub-panels of Fig.~\ref{fig:posteriorsdeltail}, but for the case where only SNeIa for $z<0.51$ are considered.}
\label{fig:posteriorsdeltailcut}
\end{figure*}

We remind the reader that, aside from the strong shape wall discussed above, the usual ``normalization wall'' remains to prevent late-time solutions to the Hubble tension~\cite{Bernal:2016gxb,Addison:2017fdm,Lemos:2018smw,Aylor:2018drw,Schoneberg:2019wmt,Knox:2019rjx,Arendse:2019hev,Efstathiou:2021ocp,Cai:2021weh,Keeley:2022ojz}. Strictly speaking, this normalization wall is only operative in the redshift range probed by BAO measurements. In the case of our specific GP reconstruction, this corresponds to $z>0.51$, since the volume-averaged \textit{DESI} BGS measurement has been excluded. It is therefore natural to ask what is the strength of the SNeIa-only ``shape wall'' for $z<0.51$. To address this question, we repeat our analysis using only $z<0.51$ SNeIa, and estimate the allowed relative variation in the $I_<$ integral, setting $z_{\max}=0.51$ in the first of Eq.~(\ref{eq:ilig}): we refer to the corresponding quantity as $\delta I_{<0.51}/I_{<0.51}^{\text{fid}}$. For the \textit{PantheonPlus}, \textit{DESY5Dvk}, and \textit{Union3} samples, we find negative central values of $\delta I_{<0.51}/I_{<0.51}^{\text{fid}}=-1.27\%$, $-3.0\%$, and $-2.05\%$ respectively. The corresponding $68\%$ ($95\%$) upper limits we find for the same samples are instead $-0.51\%$ ($0.20\%$), $-0.80\%$ ($0.80\%$), and $-1.46\%$ ($-0.80\%$) respectively. The negative central values we find in all three cases are not surprising, and reflect the trend observed in all compilations of the reconstructed dimensionless expansion rate lying above the fiducial one at low redshifts (see the three upper sub-panels of Fig.~\ref{fig:reconstructions}). If one insists on interpreting this trend within a given dark energy model at fixed $\Omega_m$, it would be associated to low-redshift values of $w>-1$, which enhance $E(z)$ and correspondingly reduce the dimensionless distance integral. Importantly, we find that none of the three cases allow for a sufficiently large positive variation in $H_0$, with the least constraining case leading to the $95\%$ limit $\delta I_{<0.51}/I_{<0.51}^{\text{fid}} \lesssim 0.80\%$. We therefore conclude that, for $z \lesssim 0.5$, the shape wall is even stronger, and prevents smooth late-time modifications from evading the normalization wall. The only remaining loophole is constituted by new physics operating at extremely low redshifts, below those probed by SNeIa. New physics at redshifts $z \ll 0.01$ has indeed been studied in several recent works and remains, to the best of our knowledge, a possible route towards solving the Hubble tension (see e.g.\ Refs.~\cite{Desmond:2019ygn,Ding:2019mmw,Desmond:2020wep,Cai:2020tpy,Cai:2021wgv,Perivolaropoulos:2021bds,Castello:2021uad,Odintsov:2022eqm,Odintsov:2022umu,Perivolaropoulos:2022khd,Oikonomou:2022tjm,Perivolaropoulos:2023iqj,Hogas:2023vim,Hogas:2023pjz,Mazurenko:2023sex,Huang:2024erq,Liu:2024vlt,Ruchika:2024ymt,Banik:2024yzi,Perivolaropoulos:2025gzo}). We stress that these models, which include for instance a possible very late transition in the gravitational constant, address the Hubble tension not by raising the ``early-time'' value of $H_0$, but rather bringing down the local value. Very-low-redshift effects of this type are not constrained by our analysis: they therefore remain a viable loophole, worthy of further investigation.

It is also worth reminding the reader that all our conclusions depend on a number of (more or less implicit) assumptions. For instance, our GP reconstruction assumes that the expansion history is sufficiently smooth. Very sharp features, with oscillation frequency below the ``redshift resolution'' of the data, cannot be captured by our reconstruction, and could potentially evade our results. Nevertheless, we deem this very unlikely. Furthermore, in interpreting all our measurements, we have assumed a spatially flat FLRW geometry, motivated by current very tight constraints on the spatial curvature parameter~\cite{Handley:2019tkm,Wang:2019yob,DiValentino:2019qzk,Efstathiou:2020wem,DiValentino:2020hov,Chudaykin:2020ghx,Vagnozzi:2020rcz,Vagnozzi:2020dfn,DiValentino:2020kpf,Yang:2021hxg,Cao:2021ldv,Dhawan:2021mel,Gonzalez:2021ojp,Dinda:2021ffa,Zuckerman:2021kgm,Bargiacchi:2021hdp,Akarsu:2021max,Liu:2022lqw,Glanville:2022xes,Wu:2022fmr,Yang:2022kho,Favale:2023lnp,Qi:2023oxv,Giare:2023ejv,Wu:2024faw,Liu:2024yib,Forconi:2025zzu,Specogna:2025ufe,Wu:2026qog,Comini:2026nsj,Giare:2026oti}. It is plausible that introducing non-zero spatial curvature could weaken some of our constraints. Similar considerations hold for the distance-duality relation, which we are implicitly assuming when comparing SNeIa and BAO measurements. Moreover, whereas our limits on $I_<$ are nonparametric, our limits on $I_>$ are model-guided, and may in principle be evaded. Another way to constrain the $z>z_{\max}$ region could be to make use of CMB geometrical information in our GP reconstruction, i.e.\ using a CMB-informed prior on $\theta_s$ effectively as a BAO measurement at $z=z_{\star}$. All these caveats are definitely worth exploring further, and we leave a dedicated study to follow-up work, which may become even more relevant as new CMB and BAO data become available~\cite{SimonsObservatory:2018koc,SimonsObservatory:2019qwx,Euclid:2024yrr}. Finally, we stress that, in translating our bounds on $\delta I/I$ into bounds on $\delta H_0/H_0$, we have assumed that $r_s^{\star}$ remains unchanged. However, even if early-time new physics modifies $r_s^{\star}$, our results still constrain the late-time contribution $(\delta H_0/H_0)_{\text{late}}=\delta I/I$, while the total relative shift in $H_0$ receives the additional, independent contribution $(\delta H_0/H_0)_{\text{early}}=-\delta r_s^\star/r_s^\star$, see Eq.~(\ref{eq:deltathetas}). In this sense, the shape wall we have identified quantifies the maximum increase in $H_0$ which can be contributed by late-time modifications to $E(z)$ at fixed $\theta_s$, regardless of whether or not early-time new physics is present as well.

Summing up, our results conclusively show that a percent-level \textit{shape wall} faces attempts to address the Hubble tension through late-time new physics. This is conceptually different from the familiar \textit{normalization wall} invoked in the familiar ``no-go theorem'' against late-time solutions, and even stronger statistically speaking. We stress that this shape wall, and more generally the conclusions of our results, prevents late-time modifications from fully addressing the Hubble tension \textit{on their own}, but does not imply that late-time new physics cannot play a role in the Hubble tension. As recently suggested, it is in fact possible that late-time modifications to the expansion history may provide a subdominant contribution towards partially reducing the Hubble tension~\cite{Vagnozzi:2023nrq}: whatever this contribution may be, it is subject to the shape wall identified here, which constrains the maximum possible increase in $H_0$ which can be contributed by the late-time new physics ingredients.

\section{Conclusions}
\label{sec:conclusions}

The well-known ``no-go theorem'' preventing late-time solutions to the Hubble tension is essentially a normalization argument, i.e.\ the fact that BAO data constrain the product $r_dH_0$. If $r_d$ is fixed, late-time modifications cannot produce the required increase in $H_0$ without worsening the fit to BAO data, whose amplitude thus sets a ``normalization wall''~\cite{Bernal:2016gxb,Addison:2017fdm,Lemos:2018smw,Aylor:2018drw,Schoneberg:2019wmt,Knox:2019rjx,Arendse:2019hev,Efstathiou:2021ocp,Cai:2021weh,Keeley:2022ojz}. However, recent work has emphasized that, independently of its absolute normalization, the shape of the expansion history itself, i.e.\ the dimensionless expansion rate $E(z)$, can play a particularly important role in the Hubble tension~\cite{Camarena:2021jlr,Zhou:2025kws,Pedrotti:2025ccw,Bansal:2026axl,Zhou:2026iar}. In this work, we quantify the strength of this ``shape wall''. We show that, if $E(z)$ is constrained by late-time observations, and the CMB acoustic scale $\theta_s$ is fixed, the allowed relative shift in $H_0$ which can be achieved by late-time modifications (i.e.\ at fixed sound horizon) is related to the relative change in the dimensionless distance integral $I \equiv \int dz/E(z)$: $\delta H_0/H_0 \simeq \delta I/I$.

Using late-time observations in a way which isolates shape information, i.e.\ unanchored SNeIa and BAO data, we reconstruct $E(z)$ nonparametrically and propagate these reconstructions to determine $\delta H_0/H_0$. We find that $\delta H_0/H_0$ is constrained to the percent level. In particular, the relative shift in $H_0$ allowed by new physics at $z \lesssim 2.3$, i.e.\ in the redshift range which is directly constrained by observations, is below the percent level at best (see the rightmost column of Tab.~\ref{tab:deltailpercentage}). Extending our estimate to new physics in the entire post-recombination era, we find that the relative shift is at the $\lesssim 2\%$ level in a conservative analysis, and does not exceed $\simeq 3.7\%$ even in the most permissive but least conservative analysis (see the rightmost column of Tab.~\ref{tab:deltaipercentage}). These figures fall significantly short of the $\simeq 9.3\%$ shift required to fully solve the Hubble tension~\cite{SPT-3G:2025bzu,H0DN:2025lyy}. Our main result is therefore the following: \textit{there is a percent-level shape wall, which prevents a fully late-time solution to the Hubble tension, and depends only marginally on the adopted SNeIa sample}. We stress that any attempt to alleviate the Hubble tension at late times will face obstructions from both the normalization wall \textit{and} the shape wall. It is important to clarify once more that our results do not imply that late-time new physics is irrelevant to the Hubble tension. Such models cannot fully solve the tension on their own, but may play a supporting role when combined, for instance, with early-time new physics which lowers the sound horizon to address the normalization wall. However, even these ``hybrid'' approaches will still be subject to the shape wall identified in this work, which will constrain the late-time contribution to the increase in $H_0$, $(\delta H_0/H_0)_{\text{late}}=\delta I/I$, whereas the early-time contribution is $(\delta H_0/H_0)_{\text{early}}=-\delta r_s^\star/r_s^\star$ [see Eq.~(\ref{eq:deltathetas})]. 

Our conclusions are especially important given that possible fiducial cosmology dependence or systematics are sometimes invoked as a possible caveat to the use of BAO measurements in the usual normalization wall ``no-go theorem''. Since we have explicitly marginalized over the BAO normalization, a redshift-independent shift in the BAO scale cannot circumvent the shape wall we have identified here (although the strength of the wall in light of SNeIa data alone is somewhat weaker). Unlike a normalization wall, evading a shape wall requires more than an absolute calibration shift.~\footnote{On this point, we clarify that the shape wall is a limit on the net (appropriately marginalized) variation of the Hubble constant, i.e.\ the range of feasible $H_0$ values reachable through changes in the shape of $E(z)$, is limited, even though the shape varies. In fact, the GP realizations allow considerable freedom in $E(z)$, and the specific individual realizations can differ appreciably.} Specifically, it requires, in order of decreasing plausibility, either new physics at extremely low redshifts ($z \ll 0.01$), redshift-dependent violations of the distance-duality relation (whether due to new physics or systematics), or highly localized low-redshift features. The first option has indeed been explored in the recent literature, and we deem it the most promising among the three, whereas the second is much more tightly constrained~\cite{Tiwari:2026pzk}. There are a number of caveats to our results, ranging from the assumption of a spatially flat FLRW Universe, to the implicitly assumed smoothness of the expansion history: while we expect our results to be stable against these assumptions, we defer a full study to future work. Nevertheless, the main message of our work is very clear: \textit{the case against late-time solutions to the Hubble tension is stronger than ever}.

\begin{acknowledgments}
\noindent We thank the referee for the constructive comments and suggestions, which helped to improve the clarity of the results. We acknowledge the use of the UniTN HPC cluster. M.A.S., D.P., and S.V.\ acknowledge support from the Istituto Nazionale di Fisica Nucleare (INFN) through the Commissione Scientifica Nazionale 4 (CSN4) Iniziativa Specifica ``Quantum Fields in Gravity, Cosmology and Black Holes'' (FLAG). This publication is based upon work from the COST Action CA21136 ``Addressing observational tensions in cosmology with systematics and fundamental physics'' (CosmoVerse), supported by COST (European Cooperation in Science and Technology). L.A.E.\ acknowledges partial financial support from the T\"{u}rkiye Bilimsel ve Teknolojik Ara\c{s}t{\i}rma Kurumu (T\"{U}B\.{I}TAK, Scientific and Technological Research Council of T\"{u}rkiye) through grant no.\ 124N627. Fig.~\ref{fig:summary} is based on a figure originally presented by William Giar\`{e} at a CosmoVerse seminar (\href{https://www.youtube.com/watch?v=3Vmj9mLkmtE}{www.youtube.com/watch?v=3Vmj9mLkmtE}). We thank him for sharing the corresponding plotting code, publicly available at \href{https://github.com/williamgiare/wgcosmo}{github.com/williamgiare/wgcosmo}, which we have adapted.
\end{acknowledgments}

\appendix

\section{Gaussian process methodology}
\label{app:gpmethodology}

We provide more technical details of the GP reconstruction methodology concisely summarized in Sec.~\ref{sec:datasets}. Broadly speaking, the methodology is divided in four steps: construction of appropriately rescaled data vectors, construction of the GP likelihood for $\widetilde D_M(z)$ and $\widetilde D_M'(z)$, joint MCMC inference of the GP hyperparameters and the BAO amplitude $r_dh$, and drawing correlated GP realizations of $E(z)$ from which the posterior distributions for $\delta I_</I_<^{\text{fid}}$ are estimated.

\subsection{Construction of dimensionless data vectors}
\label{app:construction}

We begin by converting the SNeIa and BAO measurements into measurements of dimensionless transverse comoving distances $\widetilde D_M(z)$ and dimensionless luminosity distances $\widetilde D_L(z)$. As discussed in the main text, for \textit{PantheonPlus} we construct distance moduli starting from the publicly released apparent magnitudes $m_{\text{obs},i}$ as follows:
\begin{equation}
\mu_i=m_{\text{obs},i}-M^{\text{fid}}\,,
\label{eq:mupp}
\end{equation}
where the fiducial absolute magnitude $M^{\text{fid}}=-19.25$ is associated to the reference fiducial value of $H_0^{\text{fid}} \simeq 73\,{\text{km}}/{\text{s}}/{\text{Mpc}}$, and we stress once more that we are not adopting the SH0ES calibrator likelihood or imposing a prior on $M$ or $H_0$, and that the value of $H_0^{\text{fid}}$ is arbitrary and will cancel later. For \textit{DESY5Dvk} and \textit{Union3}, the publicly release distance moduli $\mu_{\text{obs},i}$ are provided assuming a reference value $H_0^{\text{ref}}=70\,{\text{km}}/{\text{s}}/{\text{Mpc}}$, so we construct their distance moduli as follows:
\begin{equation}
\mu_i=\mu_{\text{obs},i}-5\log_{10}\left(\frac{H_0^{\text{fid}}}{H_0^{\text{ref}}}\right).
\label{eq:muunion3desy5}
\end{equation}
The distance moduli constructed in Eqs.~(\ref{eq:mupp},\ref{eq:muunion3desy5}) are all expressed on a common reference scale, whose actual value is nevertheless arbitrary and irrelevant to our analysis. We note that a similar strategy has been adopted in Ref.~\cite{Zhang:2025bmk}. Assuming the validity of the distance-duality relation, the associated dimensionless luminosity and transverse comoving distances are then the following:
\begin{equation}
\widetilde D_{L,i}=10^{(\mu_i-25)/5}\frac{H_0^{\text{fid}}}{c}\,, \qquad\widetilde D_{M,i}=\frac{\widetilde D_L^{(i)}}{1+z_i}\,.
\label{eq:tildadlitildadmi}
\end{equation}
From the above, it is trivial to see that the dependence on $H_0^{\text{fid}}$ cancels for all three SNeIa samples.

The above transformations of course need to be propagated to the covariance matrix of the new measurements. At a given redshift, the relevant Jacobian for the transformation $\mu \to \Tilde{D}_M$ is the following:
\begin{equation}
J_i \equiv \frac{\partial\widetilde D_M^{(i)}}{\partial\mu_i}=\frac{\ln 10}{5}\widetilde D_M^{(i)}\,.
\label{eq:jacobian}
\end{equation}
It follows that the covariance matrix for the SNeIa dimensionless comoving distances is given by the following: 
\begin{equation}
C_{ij}^{\widetilde D_M}=J_i C_{ij}^{\mu}J_j\,.
\label{eq:covariancetildadm}
\end{equation}
For the \textit{PantheonPlus} sample, individual redshift uncertainties $\sigma_{z_i}$ are available, and need to be propagated correctly. To do so, we add an additional diagonal term (see e.g.\ Ref.~\cite{Liu:2023agr}):
\begin{equation}
C_{ii}^{\widetilde D_M} \longrightarrow C_{ii}^{\widetilde D_M}+ \left [ \frac{\widetilde D_L^{(i)}}{(1+z_i)^2} \right ] ^2\sigma_{z_i}^2\,.
\label{eq:covariancetildadmsigmaz}
\end{equation}
This correction is not included for \textit{DESY5Dvk} and \textit{Union3}, as individual redshift uncertainties are not provided for them.

For what concerns BAO measurements of $D_M/r_d$ and $D_H/r_d$, our strategy is to transform these into dimensionless distance measurements up to an overall normalization $\alpha$, which will be inferred (in such a way as to be consistent with the SNeIa one) and marginalized over. We perform the following transformations:
\begin{equation}
\widetilde D_M^{\text{BAO}}(z_i)=\alpha \left [ \frac{D_M(z_i)}{r_d} \right ] \,, \qquad \widetilde D_H^{\text{BAO}}(z_i)=\alpha \left [ \frac{D_H(z_i)}{r_d} \right ] \,,
\label{eq:dimensionlessdistancebao}
\end{equation}
where the normalization $\alpha$ is defined as follows:
\begin{equation}
\alpha\equiv\frac{H_0r_d}{c}=r_dh\frac{100\,{\text{km}}/{\text{s}}/{\text{Mpc}}}{c}\,.
\label{eq:appalpha}
\end{equation}
We will later sample $r_dh$ jointly with the GP hyperparameters, and this implies that both the BAO data vector and covariance matrix depend on the value of $r_dh$ at each MCMC step. If we denote the starting data vector and covariance matrix by $\boldsymbol{d}_{\text{BAO}}$ and $\boldsymbol{C}_{\text{BAO}}$ respectively, it is easy to show that their rescaled counterparts are rescaled by $\alpha$ and $\alpha^2$ respectively. Since we apply the transformation to the full \textit{DESI} covariance matrix, information on the cross-covariance between transverse and line-of-sight measurements is preserved by construction. Marginalizing over $r_dh$ will therefore remove information on the BAO normalization, retaining only shape information. In other words, we are using BAO measurements as a relative distance ruler to constrain $E(z)$.

\subsection{Gaussian Process likelihood}
\label{app:likelihood}

We model $\widetilde D_M(z)$ as a Gaussian process $f(z) \equiv \widetilde D_M(z)$ constrained by SNeIa and transverse BAO measurements, and whose derivative $f'(z) \equiv \widetilde D_H(z)$ is also a Gaussian process, constrained by line-of-sight BAO measurements. We impose the physically motivated boundary conditions $f(0)=0$ and $f'(0)=1$, which capture the fact that comoving distances vanish at the present time, and the present-day dimensionless expansion rate is $1$ by definition. In practice, these boundary conditions are treated as two additional datapoints with negligible variance ($\sigma^2 \approx 0$), which we concatenate with SNeIa and BAO measurements.

Considering a joint SNeIa+BAO analysis, the full function data vector and redshift vector are therefore given by the following:
\begin{align}
\boldsymbol{Y} &= \left ( 0,\widetilde D_{M,1}^{\text{SNeIA}},\ldots,\widetilde D_{M,N_{\text{SNeIa}}}^{\text{SN}},\widetilde D_{M,1}^{\text{BAO}},\ldots,\widetilde D_{M,N_{\text{BAO}}}^{\text{BAO}} \right ) ^T\,,\\
\boldsymbol{Z} &= \left ( 0,z_1^{\text{SNeIa}},\ldots,z_{N_{\text{SNeIa}}}^{\text{SN}},z_1^{\text{BAO}},\ldots,z_{N_{\text{BAO}}}^{\text{BAO}} \right ) ^T\,,
\label{eq:functiondatavector}
\end{align}
where the notation is self-explanatory. The corresponding data vectors for the derivative observations are instead the following:
\begin{align}
\boldsymbol{Y}' &= \left ( 1,\widetilde D_{H,1}^{\text{BAO}},\ldots,\widetilde D_{H,N_{\text{BAO}}}^{\text{BAO}} \right ) ^T\,, \\
\boldsymbol{Z}' &= \left ( 0,z_1^{\text{BAO}},\ldots,z_{N_{\text{BAO}}}^{\text{BAO}} \right ) ^T\,.
\label{eq:functionderivativedatavector}
\end{align}
For SNeIa-only analyses, we obviously remove the BAO datapoints, which leaves only the boundary condition for the derivative observations.

The covariance matrices for the function ($f$) and derivative ($d$) measurements are block diagonal, and given by the following:
\begin{equation}
\boldsymbol{\Sigma}_{ff} = \begin{pmatrix}
\sigma_0^2 & \boldsymbol{0}^T & \boldsymbol{0}^T \\
\boldsymbol{0} & \boldsymbol C_{\text{SN}}^{\widetilde D_M} & \boldsymbol{0} \\
\boldsymbol{0} & \boldsymbol{0} & \boldsymbol C_{MM}^{\text{BAO}}
\end{pmatrix}\,,
\qquad
\boldsymbol{\Sigma}_{dd} = \begin{pmatrix}
\sigma_0^{\prime\,2} & \boldsymbol{0}^T \\
\boldsymbol{0} & \boldsymbol C_{HH}^{\text{BAO}}
\end{pmatrix}\,,
\label{eq:functionderivativecovariances}
\end{equation}
where the BAO covariance blocks have been rescaled by $\alpha^2$, as discussed earlier. Similarly, the $f$-$d$ cross-covariance is given by the following:~\footnote{We modify the \texttt{GaPP} Python library to include this cross term.}
\begin{equation}
\boldsymbol{\Sigma}_{fd}
=
\begin{pmatrix}
0 & \boldsymbol{0}^T \\
\boldsymbol{0} & \boldsymbol{0} \\
\boldsymbol{0} & \boldsymbol C_{MH}^{\text{BAO}}
\end{pmatrix}\,.
\label{eq:functionderivativecrosscovariance}
\end{equation}
Note that we set the SNeIa-BAO cross-covariances to zero, as the two datasets are completely independent.

We assume that any finite collection of points of the function $f(z)$ follows a multivariate Gaussian distribution with mean $\mu(z)=0$ the following covariance:
\begin{equation}
k(z,z')=\sigma_f^2 \exp \left [ -\frac{(z-z')^2}{2\ell^2} \right ] \,.
\label{eq:squaredexponential}
\end{equation}
In other words, we are adopting a zero-mean GP prior with a squared-exponential kernel.

The kernel covariance between function and derivative observations, and between derivative observations themselves, are obtained as follows:
\begin{align}
k_{fd}(z,z') &\equiv \frac{\partial k(z,z')}{\partial z'} = \frac{z-z'}{\ell^2}k(z,z')\,, \\
k_{dd}(z,z') &\equiv \frac{\partial^2 k(z,z')}{\partial z\,\partial z'} = \left [ \frac{1}{\ell^2} -\frac{(z-z')^2}{\ell^4} \right ] k(z,z')\,,
\label{eq:fdkernels}
\end{align}
with kernel matrices defined as follows:
\begin{align}
[\boldsymbol K_{ff}]_{ij} &= k(Z_i,Z_j)\,, \nonumber\\
[\boldsymbol K_{fd}]_{ij} &= k_{fd}(Z_i,Z_j')\,, \nonumber\\
[\boldsymbol K_{dd}]_{ij} &= k_{dd}(Z_i',Z_j')\,.
\label{eq:kernelblock}
\end{align}
The resulting joint kernel and observational covariance matrices are then given as follows:
\begin{equation}
\boldsymbol K= \begin{pmatrix}
\boldsymbol K_{ff} & \boldsymbol K_{fd}\\
\boldsymbol K_{fd}^T & \boldsymbol K_{dd}
\end{pmatrix}\,,
\qquad
\boldsymbol\Sigma_{\text{obs}}= \begin{pmatrix}
\boldsymbol\Sigma_{ff} & \boldsymbol\Sigma_{fd}\\
\boldsymbol\Sigma_{fd}^T & \boldsymbol\Sigma_{dd}
\end{pmatrix}\,.
\label{eq:kerneldatacovariance}
\end{equation}
The total covariance matrix entering the likelihood is then given by the following:
\begin{equation}
\boldsymbol C_{\text{tot}} = \boldsymbol K+\boldsymbol\Sigma_{\text{obs}}\,.
\label{eq:totalcovariance}
\end{equation}
If we define the complete data vector as $\boldsymbol y=(\boldsymbol Y,\boldsymbol Y')^T$, the GP log-likelihood can then be written as follows (up to a constant factor, irrelevant for our analysis, of $N\ln (2\pi)/2$, where $N$ is the length of the data vector):
\begin{equation}
\ln\mathcal L(\sigma_f,\ell,r_dh)=-\frac{1}{2}\boldsymbol y^T\boldsymbol C_{\text{tot}}^{-1}\boldsymbol y -\frac{1}{2}\ln\det\boldsymbol C_{\text{tot}}\,.
\label{eq:gplikelihood}
\end{equation}
We stress that both $\boldsymbol y$ and $\boldsymbol C_{\text{tot}}$ depend on $r_dh$ through the rescaling of the BAO data and the respective covariance matrices. For the SNeIa-only analyses, there is obviously no dependence on $r_dh$, so the log-likelihood in Eq.~(\ref{eq:gplikelihood}) depends solely on $\sigma_f$ and $\ell$.

\subsection{MCMC inference of GP hyperparameters and BAO amplitude}
\label{app:mcmc}

\begin{figure}[!htpb]
\centering
\includegraphics[width=1.0\linewidth]{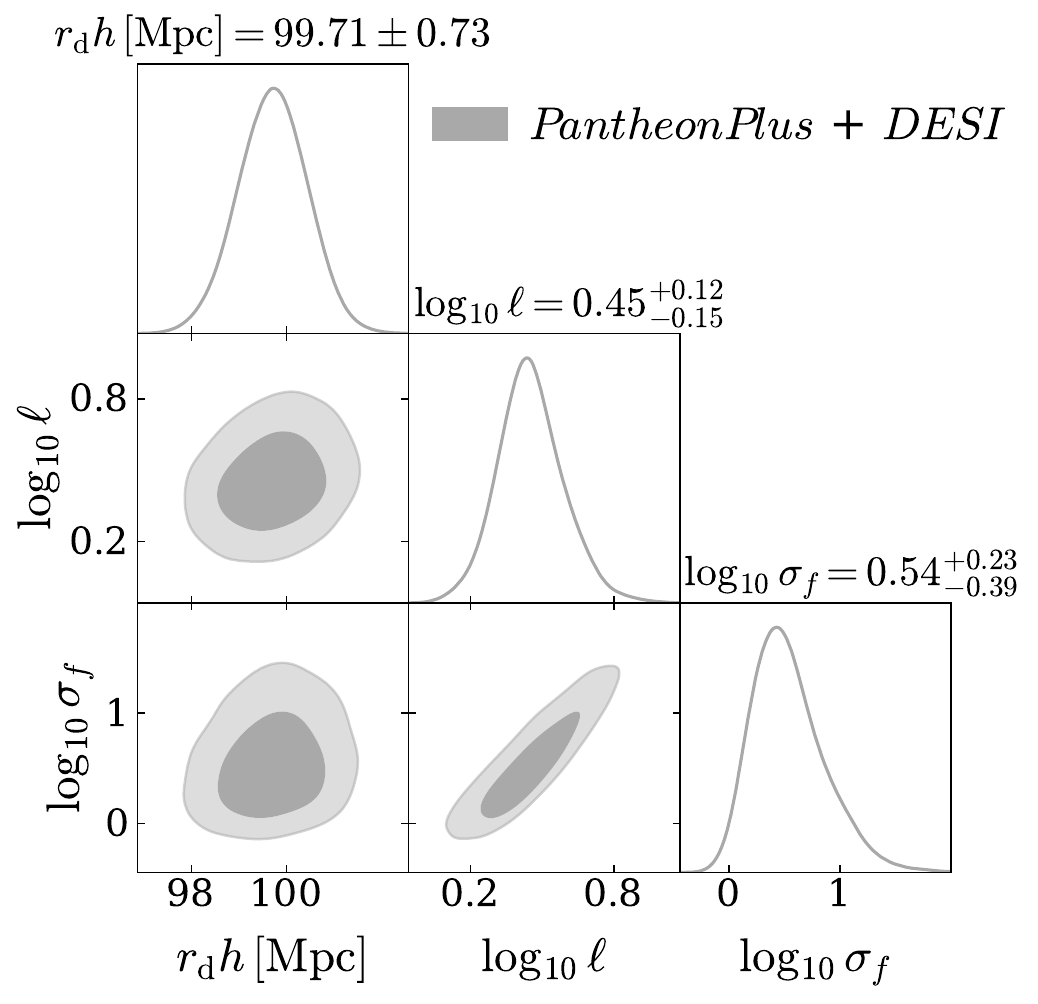}
\caption{Marginalised posteriors and 2D contours (68\% and 95\% CL) of $r_dh$ and GP hyperparameters $\log_{10}\ell$ and $\log_{10}\sigma_f$ for the dataset combination \textit{PantheonPlus} + \textit{DESI}. The priors are $r_{d}h\in[70,150]$, $\log_{10}\ell\in[-3,2]$ and $\log_{10}\sigma_{f}\in[-5,2]$. The values above each panel are the mean and $1\sigma$ limits.}
\label{fig:rd_h_posteriors}
\end{figure}

We implement the above likelihood in the \texttt{MontePython} cosmological MCMC sampler~\cite{Audren:2012wb,Brinckmann:2018cvx} to sample the joint posterior of the parameter vector $\boldsymbol{\Theta} = \{\log_{10}\sigma_f,\log_{10}\ell,r_dh\}$, which reduces to $\{\log_{10}\sigma_f,\log_{10}\ell\}$ for our SNeIa-only analyses. We set wide, flat priors on all three parameters for all our analyses, and verify that the posteriors do not depend on the choice of lower and upper prior boundaries (see Fig.~\ref{fig:rd_h_posteriors} for the marginalised posteriors from the \textit{PantheonPlus} + \textit{DESI} combination). The choice of sampling $\log_{10}\sigma_f$ and $\log_{10}\ell$ rather than $\sigma_f$ and $\ell$ allows us to efficiently explore a wide range of scales, given that the best-fit values of these parameters are not known a priori, since we do not train them. As for $r_dh$, it is treated as a nuisance parameter which controls the BAO amplitude, and no external information on $r_d$ or $H_0$ is provided.

We sample the posterior distribution of $\boldsymbol{\Theta}$ using MCMC methods. We discard the first $30\%$ of points of each chain as burn-in, and assess the convergence of the chains using the Gelman-Rubin diagnostic $R-1$~\cite{Gelman:1992zz}, requiring $R-1<0.01$ for all three parameters for our chains to be considered converged. We stress that our later analysis does \textit{not} fix the GP hyperparameters or $r_dh$ to the posterior means or best-fit values. We instead use the full converged chains to propagate the uncertainties on these parameters onto the reconstructed dimensionless expansion histories, as we will discuss shortly.

\subsection{Dimensionless expansion history reconstruction}
\label{app:ezreconstruction}

For each of the six dataset combinations, we draw $N_{\text{chain}}$ samples $\boldsymbol{\Theta}_k$ from our MCMC chains. For every sample, we fix the GP hyperparameters and (for the SNeIa+BAO analyses) the BAO normalization to their sampled values, while conditioning the GP on the complete dataset. We then evaluate the posterior of $\widetilde D_M'(z)$ on the uniform redshift grid $z^{\star} = (z_1^\star, \ldots, z_{n^\star}^\star)$, which extends from $z=0$ to the relevant $z_{\max}$. The reconstruction is therefore \textit{not} extrapolated beyond the redshift range directly covered by the data. Denoting by $\boldsymbol K_{y,k}^{\star}$ the cross-covariance between the derivative GP evaluated on $z^\star$ and the complete set of function and derivative observations, the posterior mean and covariance of $\widetilde D_M'(z)$ are then given by the following:
\begin{align}
\boldsymbol\mu_k^\star &= \boldsymbol K_{y,k}^\star \boldsymbol C_{\text{tot},k}^{-1} \boldsymbol y_k\,, \nonumber\\
\boldsymbol C_k^\star &= \boldsymbol K_{dd,k}^{\star\star}
-\boldsymbol K_{y,k}^\star \boldsymbol C_{\text{tot},k}^{-1} \left ( \boldsymbol K_{y,k}^\star \right ) ^T\,,
\label{eq:posteriorgp}
\end{align}
where $\boldsymbol K_{dd,k}^{\star\star}$ is the derivative-derivative kernel matrix evaluated on the reconstruction grid, whereas $\boldsymbol C_{\text{tot},k}$ and $\boldsymbol y_k$ are the total covariance matrix and data vector evaluated at the sampled parameter values $\boldsymbol\Theta_k$ (recall that these depend explicitly on $r_dh$).

For each MCMC sample, once $\boldsymbol\mu_k^\star$ and $\boldsymbol C_k^\star$ are evaluated, we draw $p$ complete realizations from the corresponding GP posterior. We first perform the Cholesky decomposition as follows:
\begin{equation}
\boldsymbol C_k^\star=\boldsymbol L_k^\star(\boldsymbol L_k^\star)^T\,,
\label{eq:cholesky}
\end{equation}
where $\boldsymbol L_k^\star$ is lower triangular. The $m$-th realization is then constructed as follows:
\begin{equation}
\widetilde{\boldsymbol D}_{M,km}^{\prime\,\star}=\boldsymbol\mu_k^\star+\boldsymbol L_k^\star\boldsymbol\xi_{km}\,, \quad \boldsymbol\xi_{km} \sim \mathcal N(\boldsymbol 0,\boldsymbol I)\,.
\label{eq:draw}
\end{equation}
This ensures that each realization has the conditional mean and covariance given by Eq.~(\ref{eq:posteriorgp}), and preserves the correlations between different points on the reconstruction grid, which would be lost if we had drawn independently from the marginal posterior at each redshifts. We only keep realizations which satisfy the physical condition $\widetilde D_{M,km}'(z_j^\star)>0$, as well as the following additional smoothness condition:
\begin{equation}
\left \vert \frac{\widetilde D_{M,km}'(z_{j+1}^\star)}{\widetilde D_{M,km}'(z_j^\star)} \right \vert <3\,,
\label{eq:smoothness}
\end{equation}
for each pair of consecutive grid points. For each realization which meets these conditions, we then construct the dimensionless expansion history as follows:
\begin{equation}
E_{km}(z) = \frac{1}{\widetilde D_{M,km}'(z)}\,.
\label{eq:ezsmooth}
\end{equation}
We first perform the above inversion, and only then average the resulting ensemble of $E_{km}(z)$ at each redshift, to obtain the posterior mean reconstruction shown in Fig.~\ref{fig:reconstructions} and Fig.~\ref{fig:reconstructionscut}. In other words, what we compute is $\langle E(z)\rangle \equiv \langle
1/\widetilde D_M'(z) \rangle$, rather than $1/\langle \widetilde D_M'(z) \rangle$. The pointwise credible limits are obtained from the corresponding quantiles of the ensemble of $E_{km}(z)$ values at each grid point. Finally, for each realization, we compute $I_{<,km}$ as follows:
\begin{equation}
I_{<,km}=\int_0^{z_{\max}}
\frac{dz}{E_{km}(z)}=\int_0^{z_{\max}} \widetilde D_{M,km}'(z)\,dz
\label{eq:ilsmooth}
\end{equation}
and therefore the allowed relative variation in $I_<$, given by the following:
\begin{equation}
\frac{\delta I_{<,km}}{I_<^{\text{fid}}}=\frac{I_{<,km}-I_<^{\text{fid}}}{I_<^{\text{fid}}}\,.
\label{eq:deltailreconstruction}
\end{equation}
Each realization therefore maps onto a value of $\delta I_{<,km}/I_<^{\text{fid}}$, which is effectively treated as a derived parameters, and whose distribution is our estimate for the posterior distribution of $\delta I_</I_<^{\text{fid}}$. We stress that our procedure keeps all information on the correlation between different points in the reconstruction, and avoids the otherwise incorrect treatment of taking GP limits as complete reconstructed dimensionless expansion histories (see Sec.~\ref{sec:discussion} for further discussions on this point).

We emphasize that only in the pure SNeIa-only cases do the imposed conditions lead to discarded realizations,~\footnote{We stress that rejecting realizations that violate these conditions is not equivalent to imposing the same conditions as hard cuts on the prior. We adopt the former approach for practical reasons. Developing a more robust method for imposing such constraints on GP realizations is left to future work.} with violations appearing solely above $z \sim 1$ for \textit{DESY5Dvk}~(D5) and $z \sim 1.5$ for \textit{PantheonPlus}~(PP) and \textit{Union3}~(U3). The fractions of discarded realizations are approximately $0.5\%\, (2\%)$ for D5, $8\%\, (14\%)$ for PP, and $15\%\, (20\%)$ for U3, for the smoothness (physical) condition. For all other dataset combinations, there are no rejections, since every realization satisfy both conditions.

As a further consistency check, we also repeat our reconstruction procedure making use of the Mat\'{e}rn-$7/2$ and Mat\'{e}rn-$9/2$ kernels in place of the squared-exponential one. We find our results to be largely unchanged and stable against the change of kernel, and therefore do not show the results here.

\section{Alternative node-based methodology}
\label{app:alternativegpmethodology}

\begin{table}[!t]
\footnotesize
\resizebox{0.9\columnwidth}{!}{
\begin{tabular}{|c?c|c|}
\hline
\multirow{2}{*}{\textbf{$N_{\text{node}}$}} & \multicolumn{2}{c|}{$\delta I_</I_<^{\text{fid}}$} \\
\cline{2-3}
& Credible interval $(\%)$ & Upper $68\%$ $(95\%)$ \\
\hline\hline
$5$ & $0.52^{+0.36(+0.72)}_{-0.36(-0.72)}$ & $0.88 \quad (1.24)\%$ \\ \hline
$7$ & $0.05^{+0.50(+0.99)}_{-0.50(-0.99)}$ & $0.55 \quad (1.04)\%$ \\ \hline
$10$ & $0.27^{+0.69(+1.37)}_{-0.69(-1.37)}$ & $0.96 \quad (1.64)\%$ \\ \hline
$12$ & $0.24^{+0.69(+1.38)}_{-0.69(-1.38)}$ & $0.93 \quad (1.62)\%$ \\ \hline
$15$ & $0.22^{+0.65(+1.29)}_{-0.65(-1.29)}$ & $0.87 \quad (1.51)\%$ \\ \hline
$20$ & $0.19^{+0.61(+1.23)}_{-0.61(-1.23)}$ & $0.80 \quad (1.42)\%$ \\ \hline
\end{tabular}}
\caption{Constraints on $\delta I_</I_<^{\text{fid}}$ obtained from the \textit{PantheonPlus}+\textit{DESI} dataset combination using our alternative node-based methodology, for different numbers of nodes $N_{\text{node}}$, as reported in the first column. We report both $68\%$ and $95\%$ credible intervals (second column), as well as the corresponding $68\%$ and $95\%$ upper limits (third column). The results are in excellent agreement with those obtained from our baseline GP reconstruction methodology (see the fourth row in Tab.~\ref{tab:deltailpercentage}).}
\label{tab:alternativegp}
\end{table}

As a consistency check, we repeat our \textit{PantheonPlus}+\textit{DESI} analysis using an independent node-based reconstruction methodology. This approach was recently used in Refs.~\cite{Gerardi:2019obr,Akarsu:2026anp}, and makes use of a GP kernel, although not in the usual GP regression sense. We instead place a set of $N_{\text{node}}$ nodes at fixed, equispaced redshifts $z_i$, between $0$ and $z_{\max}$. The node amplitudes $E_i \equiv E(z_i)$ are then treated as free parameters which we constrain using the data. In between the nodes, we construct a continuous expansion history using the squared-exponential kernel given by Eq.~(\ref{eq:squaredexponential}), where we set $\sigma_f=1$. In other words, we are using GPs only as an interpolation method, whereas the uncertainty in the reconstructed dimensionless expansion history is propagated from the uncertainty in the node amplitudes and GP hyperparameters. We refer the reader to Sec.~IIIA of Ref.~\cite{Akarsu:2026anp} for a more complete discussion of the method.

Following Ref.~\cite{Akarsu:2026anp}, we sample the joint posterior of the node amplitudes $E_i$ and all nuisance parameters using nested sampling, as implemented via the \texttt{dynesty} Python library within the \texttt{SimpleMC} sampler. For each sample, we then construct the interpolated $E(z)$, and calculate $I_<$ and $\delta I_</I_<^{\text{fid}}$ as we did earlier. We repeat the reconstruction procedure $N_{\text{node}}=5,7,10,12,15$, and $20$ nodes, noting that increasing the number of nodes increases the number of parameters to be marginalized on. Since we do not aim to perform a complete comparison for all dataset combinations, we limit our analysis to the \textit{PantheonPlus}+\textit{DESI} dataset combination.

The resulting constraints on $\delta I_</I_<^{\text{fid}}$ are reported in Tab.~\ref{tab:alternativegp}. In a broad brush we see that, for all choices of $N_{\text{node}}$, the $68\%$ and $95\%$ upper limits on $\delta I_</I_<^{\text{fid}}$ lie within $\approx 0.55\%$-$0.95\%$ and $\approx 1.04\%$-$1.65\%$ respectively. This is in excellent agreement with the corresponding limits of $0.64\%$ and $1.5\%$ respectively (see Tab.~\ref{tab:deltailpercentage}), obtained from our baseline GP reconstruction methodology. This lends further support to the robustness of the main results of this work, i.e.\ the presence of a strong, percent-level shape wall preventing a fully late-time solution to the Hubble tension.

\section{Dependency test on $\Omega_{\rm m}^{\rm fid}$}
\label{app:omegamdependence}

For our $\delta I_{<}/I_{<}^{\rm fid}$ constraints presented in Sec.~\ref{sec:results}, the dependence on $\Omega_{\rm m}$ enters solely through the fiducial value $I_{<}^{\rm fid}$, see Eq.~\eqref{eq:ilfid}. Throughout the manuscript we adopt the baseline value $\Omega_{\rm m}^{\rm fid}=0.31$, based on the precise measurement by Planck 2018~\cite{Planck:2018vyg}, although this estimate of the matter content includes information from the late-time CMB lensing effect~\cite{Planck:2018lbu,Lewis:2006fu}. A conservative choice is to test how the posterior changes when $\Omega_{\rm m}^{\rm fid}$ is varied over the range allowed by the CMB TT+lowE data alone, $\Omega_{\rm m}^{\rm fid}\in[0.26,0.34]$. Since $I_<$ is fixed by the data, we expect that changing the fiducial cosmology only produces a shift on the posterior.

\begin{figure}[htpb!]
\centering
\includegraphics[width=1.0\linewidth]{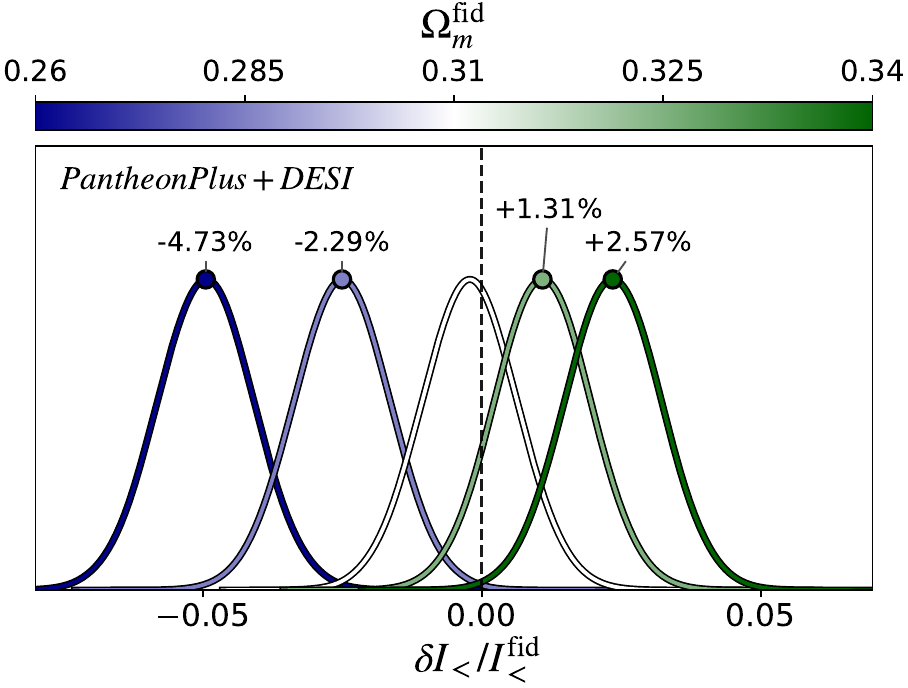}
    \caption{Posterior of $\delta I_{<}/I_{<}^{\rm fid}$ using the
    \textit{PantheonPlus} + \textit{DESI} dataset combination for different fiducial values of
    $\Omega_{\rm m}^{\rm fid}$ displayed in the color bar. The dashed line marks $\delta I_<=0$, i.e.\ $I_<=I_<^{\rm fid}\left(\Omega_{\rm m}^{\rm fid}\right)$. The labels quantify the shift of the mean in percent, with respect
    to the baseline case displayed in white, $\Omega_{\rm m}^{\rm fid}=0.31$.}
    \label{fig:Om_test}
\end{figure}

 Figure~\ref{fig:Om_test} shows the result for the \textit{PantheonPlus} + \textit{DESI} dataset combination, where the width of the posterior is essentially unchanged, while the mean moves from $-4.7\%$ for the lowest value, $\Omega_{\rm m}^{\rm fid}=0.26$, to $+2.6\%$ for the largest, $0.34$, with respect to our baseline. If one propagates this behavior to our conservative analysis discussed in Sec.~\ref{sec:discussion}, even assuming the posterior with the most shifted mean to the right, the result still falls short and fails to resolve the Hubble tension.

\clearpage

\bibliographystyle{apsrev4-1}
\bibliography{shapehubble}

\end{document}